\documentclass[%
superscriptaddress,
preprint,
amsmath,amssymb,
]{revtex4-2}

\usepackage{multirow}
\usepackage{graphicx}
\usepackage{subfigure}
\usepackage{dcolumn}
\usepackage{bm}
\usepackage{hyperref}
\usepackage{color}

\begin{document}

\preprint{APS/123-QED}

\title{Anomalous thermal Hall effect and anomalous Nernst effect of CsV$_{3}$Sb$_{5}$}

\author{Xuebo Zhou}
\affiliation{School of Physics, Harbin Institute of Technology, Harbin 150001, China}
\affiliation{Beijing National Laboratory for Condensed Matter Physics and Institute of Physics, Chinese Academy of Sciences, Beijing 100190, China}

\author{Hongxiong Liu}
\affiliation{Beijing National Laboratory for Condensed Matter Physics and Institute of Physics, Chinese Academy of Sciences, Beijing 100190, China}

\author{Wei Wu}
\affiliation{Beijing National Laboratory for Condensed Matter Physics and Institute of Physics, Chinese Academy of Sciences, Beijing 100190, China}

\author{Kun Jiang}
\affiliation{Beijing National Laboratory for Condensed Matter Physics and Institute of Physics, Chinese Academy of Sciences, Beijing 100190, China}
\affiliation{School of Physical Sciences, University of Chinese Academy of Sciences, Beijing 100190, China}

\author{Youguo Shi}
\affiliation{Beijing National Laboratory for Condensed Matter Physics and Institute of Physics, Chinese Academy of Sciences, Beijing 100190, China}

\author{Zheng Li}
\email{lizheng@iphy.ac.cn}
\affiliation{Beijing National Laboratory for Condensed Matter Physics and Institute of Physics, Chinese Academy of Sciences, Beijing 100190, China}
\affiliation{School of Physical Sciences, University of Chinese Academy of Sciences, Beijing 100190, China}

\author{Yu Sui}
\email{suiyu@hit.edu.cn}
\affiliation{School of Physics, Harbin Institute of Technology, Harbin 150001, China}
\affiliation{Laboratory for Space Environment and Physical Sciences, Harbin Institute of Technology, Harbin 150001, China}

\author{Jiangping Hu}
\affiliation{Beijing National Laboratory for Condensed Matter Physics and Institute of Physics, Chinese Academy of Sciences, Beijing 100190, China}
\affiliation{School of Physical Sciences, University of Chinese Academy of Sciences, Beijing 100190, China}

\author{Jianlin Luo}
\email{jlluo@iphy.ac.cn}
\affiliation{Beijing National Laboratory for Condensed Matter Physics and Institute of Physics, Chinese Academy of Sciences, Beijing 100190, China}
\affiliation{School of Physical Sciences, University of Chinese Academy of Sciences, Beijing 100190, China}
\affiliation{Songshan Lake Materials Laboratory, Dongguan 523808, China}


\begin{abstract}
Motived by time-reversal symmetry breaking and giant anomalous Hall effect in kagome superconductor \textit{A}V$_3$Sb$_5$ (\textit{A} = Cs, K, Rb), we carried out the thermal transport measurements on CsV$_3$Sb$_5$. In addition to the anomalous Hall effect, the anomalous Nernst effect and the anomalous thermal Hall effect emerge. Interestingly, the longitudinal thermal conductivity $\kappa_{xx}$ largely deviates from the electronic contribution obtained from the longitudinal conductivity $\sigma_{xx}$ by the Wiedemann-Franz law. In contrast, the thermal Hall conductivity $\kappa_{xy}$ is roughly consistent with the Wiedemann-Franz law from electronic contribution. All these results indicate the large phonon contribution in the longitudinal thermal conductivity. Moreover, the thermal Hall conductivity is also slightly greater than the theoretical electronic contribution, indicating other charge neutral contributions.
More than that, the Nernst coefficient and Hall resistivity show the multi-band behavior with possible additional contribution from Berry curvature at the low fields.



\end{abstract}


\maketitle


\section{Introduction}
The newly discovered kagome topological metal $A$V$_3$Sb$_5$ ($A$ = K, Rb, Cs)  has  been found to be the first quasi-two-dimensional  kagome superconductor, which becomes another platform to investigate the interplay of topology, electron correlation effects, and superconductivity \cite{Ortiz2019New,Ortiz2020Cs,Ortiz2021KVSb,Lei2021Rb,jiang2021kagome}. However, whether this superconductor is unconventional owing to electron-electron correlation or conventional because of electron-phonon coupling is still under debate. From the unconventional  side, both the V-shape tunneling  density of states \cite{zhao2021cascade} and the zero-temperature residual thermal conductivity \cite{zhao2021nodal} have been reported. And recent low-temperature scanning tunneling microscopy (STM) resolved an unconventional pair density wave \cite{chen2021roton}.
In contrast, nuclear magnetic resonance (NMR) measurements clearly reveal the spin singlet nature  and point to a conventional \textit{s}-wave superconductor from the Hebel-Slichter coherence peak \cite{Mu2021CPL}. This nodeless property is also consistent with penetration depth measurement \cite{duan2021nodeless} and impurity scattering feature from STM \cite{xu2021multiband}.  Additionally, high resolution angle-resolved photoemission spectroscopy (ARPES) finds the weakly correlated nature of $A$V$_3$Sb$_5$ and remarkable electron-phonon coupling \cite{Luo2022NC}.

\begin{figure*}
	\centering
	\includegraphics[width=0.8\textwidth,clip]{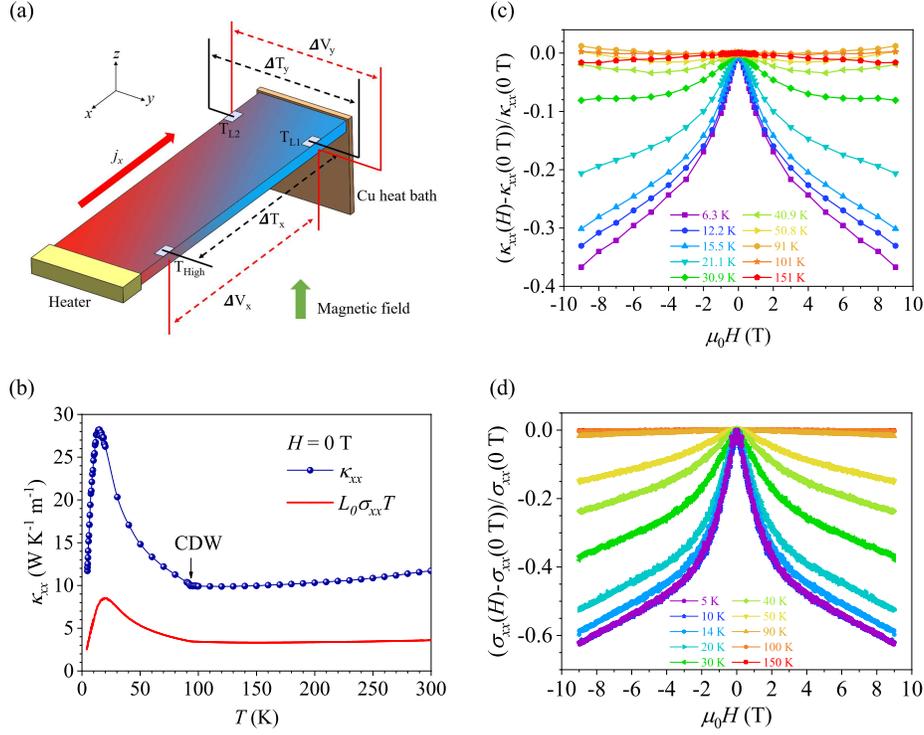}
	\caption{\label{fig:conductivity} (a) Schematic diagram of experimental setup. (b) Temperature dependence of longitudinal thermal conductivity $\kappa_{xx}$. The red curve is the electronic thermal conductivity calculated according to the WF law. Its value is smaller than the total thermal conductivity, which indicates that the phonons play an important role in thermal transportation. (c) Magnetic field dependence of longitudinal thermal conductivity $\kappa_{xx}$. (d) Magnetic field dependence of electrical conductivity $\sigma_{xx}$. Above $40$ K, the dependence of $\sigma_{xx}$ on magnetic field is weak, while below $40$ K, $\sigma_{xx}$ decreases rapidly with the field increasing.}
\end{figure*}

Besides superconductivity, the charge density wave (CDW) order in this nonmagnetic $A$V$_3$Sb$_5$ seems to be more unconventional. Several time-reversal symmetry-breaking signals have been found \cite{jiang2020discovery,Shumiya2021PRB,mielke2021timereversal}, especially the internal magnetic field induced relaxation rates in muon spin rotation ($\mu$SR) \cite{yu2021evidence}. $A$V$_3$Sb$_5$ also shows a giant anomalous Hall effect (AHE), which can be attributed to the enhanced skew scattering in the CDW state and the large Berry curvature from the kagome lattice and time-reversal symmetry breaking \cite{Yang2020Hall,yu2021concurrence,Nagaosa2010REVIEWSOFMODERNPHYSICS,XiaoDi2010REVIEWSOFMODERNPHYSICS}. This AHE is concurrent with the CDW, indicating a strong correlation between CDW and AHE.
As the thermoelectric counterpart and thermal counterpart of anomalous Hall effect, anomalous Nernst effect (ANE), and anomalous thermal Hall effect (ATHE) also play an important role in understanding quantum materials \cite{Li2020MRS,XiaoDi2006PHYSICALREVIEWLETTERS,OnodaS2008PhysicalReviewB,ShiomiY2010PhysicalReviewB}.
When the longitudinal heat flow in a magnetic field, the thermal Hall effect generates a transverse temperature gradient and the Nernst effect generates a transverse electric field, as illustrated in Fig. \ref{fig:conductivity} (a).
Normally, the Wiedemann-Franz (WF) law establishes a direct connection between electronic transport and electronic thermal transport \cite{LiXiaokang2017PhysRevLett,ZhuZengwei2020}. However, unlike the charge transport, other charge neutral excitations (e.g., phonons) can also contribute to the thermal transport. And the thermal Hall effect has been applied to probe topologically nontrivial excitations in insulators \cite{GrissonnancheG2019NATURE,Hirschberger2015SCIENCE,OnoseY2010SCIENCE}.
Hence, thermal transport provides a unique way toward identifying excitations and their topological properties.

In this work, we conducted extensive researches on the thermal transport properties of CsV$_3$Sb$_5$, especially the anomalous thermal Hall and the anomalous Nernst effect. Interestingly, we found that the longitudinal thermal conductivity $\kappa_{xx}$ largely deviates the electronic thermal conductivity contribution down to $5$ K. It indicates the charge neutral excitation contribution to the thermal transport. However, the thermal Hall conductivity is more or less consistent with the Hall conductivity according to WF law. Since phonons normally do not contribute to the thermal Hall, $\kappa_{xx}$ must contain a large phonon part down to low temperature.
At low temperature, the thermal Hall conductivity is still slightly higher than the theoretical value $L_{0}\sigma_{xy}T$, indicating that the thermal Hall effect may have phonon-drag mechanism or other electrically neutral low energy excitation \cite{Strohm2005prl,Sheng2006prl,Kagan2008prl,Mori2014prl}. Additionally, we found the Nernst coefficient reaches a maximum value when the Hall resistivity is zero in the high magnetic field. This is the characteristic of ambipolar transport  due to the multiband nature of CsV$_3$Sb$_5$. However, this relationship no longer holds as the magnetic field decreases. Hence, there are other factors that dominate the transport properties at the low magnetic field. The Berry curvature may contribute to the anomalous transverse response in the low field in addition to the multiband effect \cite{ZhuZengwei2020, Guolin2021arxiv}.

\begin{figure*}
	\centering
	\includegraphics[width=0.95\textwidth,clip]{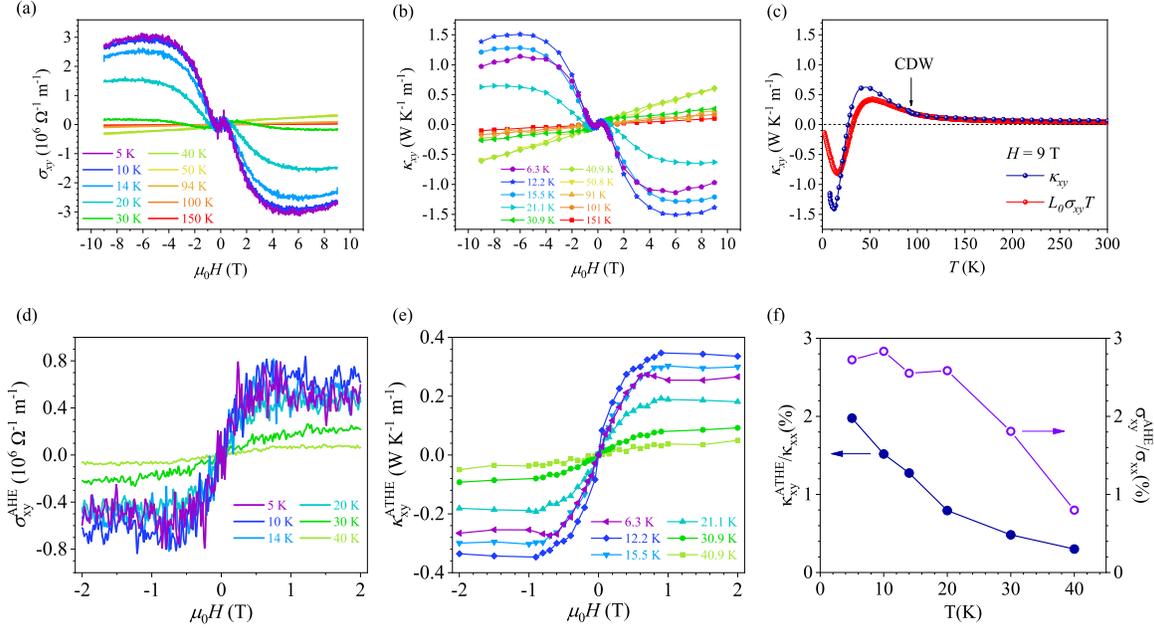}
	\caption{\label{fig:Hall} (a) Magnetic field dependence of electrical Hall conductivity $\sigma_{xy}$. An appreciable antisymmetric side ``S'' line pattern appears in the low magnetic field region below $40$ K. (b) Magnetic field dependence of thermal Hall conductivity $\kappa_{xy}$. An antisymmetric side ``S'' line pattern appears in the low magnetic field region below $40$ K. (c) Temperature dependence of thermal Hall conductivity $\kappa_{xy}$. The red curve is the electronic thermal Hall conductivity calculated according to the WF law. (d) The extracted anomalous Hall conductivity at various temperatures. (e) The extracted anomalous thermal Hall conductivity at various temperatures. (f) Temperature dependence of anomalous thermal Hall ratio and anomalous Hall ratio in $2$ T magnetic field}.
\end{figure*}

\section{Experimental details}
High-quality CsV$_3$Sb$_5$ single crystals are prepared by a two-step self-flux method \cite{Ortiz2019New}. The thermal and electrical transport measurements were performed in a 9-T Quantum Design physical property measurement system.
Schematic diagram of the experimental setup is shown in Fig. $1$ (a). The sample is processed into a square piece.
A resistance heater is used to generate a heat current $j_x = P/(Wd)$ in the plane, where $P$, $W$, and $d$ are the heater power and the width and the thickness of the sample, respectively.
The longitudinal voltage $\Delta V_x$ and the transverse voltage $\Delta V_y$ are measured by a Keithley 2182A nanovoltmeter. The longitudinal and transverse temperature gradients $\Delta T_x/l $ and $\Delta T_y/w$ were measured with field-calibrated chromel-AuFe$_{0.07\%}$ thermocouples, where $l$ is the distance between the thermal contacts for $T_{\rm High}$ and $T_{\rm L1}$ and $w$ is the distance between the thermal contacts for $T_{\rm L1}$ and $T_{\rm L2}$.
The longitudinal thermal conductivity is measured by $\kappa_{xx} = j_x l/\Delta T_x$ and the thermal Hall conductivity is measured by $\kappa_{xy} = - \kappa_{xx} [(\Delta T_y l)/(\Delta T_x w)]$.
In order to eliminate the misalignment effect of thermocouple, $\Delta V_y(H) = [\Delta V_y(+H)-\Delta V_y(-H)]/2$ and $\Delta T_y(H) = [\Delta T_y(+H)-\Delta T_y(-H)]/2$ are obtained by subtracting the measurement results of the positive and negative fields, where $H$ is the magnetic field perpendicular to the plane. Nernst coefficient $N$ is calculated by $N = (\Delta V_y l)/(\Delta T_x w)$ and it's sign is defined by using the vortex convention.

\section{Result and discussions}

\begin{figure}
	\centering
	\includegraphics[width=8.5cm]{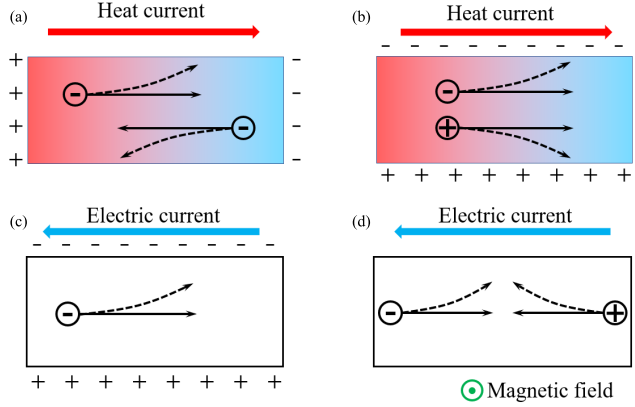}
	\caption{\label{fig:schematic} [(a), (b)] Schematic diagram of the Nernst effect for a single-band metal and an ambipolar metal, respectively. In the single-band metal, the transverse charge current produced by the thermal gradient and electric field in a magnetic field will gather at both ends, so it will not produce a transverse electric field. However, in the ambipolar metal, the holes and the electrons will move toward the cold end under the drive of the thermal gradient and accumulate to both transverse ends in the magnetic field to produce a limited electric field. [(c), (d)] Schematic diagram of the Hall effect for a single-band metal and an ambipolar metal, respectively. In the ambipolar metal, the holes and the electrons moving in opposite direction will accumulate to the same transverse end under the magnetic field, resulting in cancellation of the transverse electric field. It means that the Nernst coefficient will appear a maximum value when the Hall effect is completely offset.}
\end{figure}

\begin{figure*}
	\centering
	\includegraphics[width=0.95\textwidth,clip]{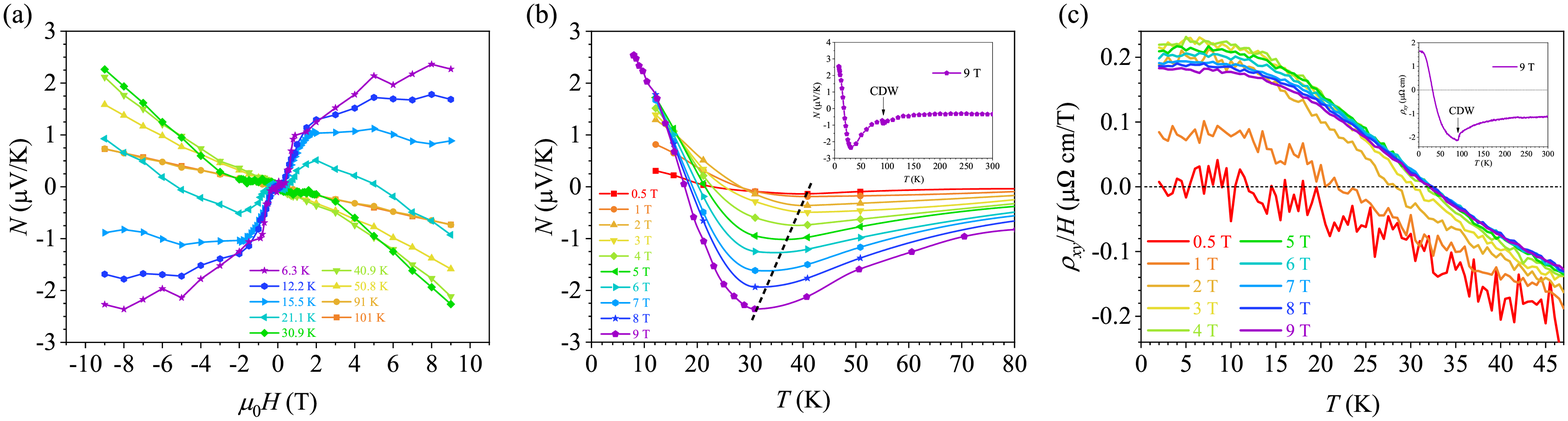}
	\caption{\label{fig:Nernst}(a) Magnetic field dependence of Nernst coefficient. The Nernst coefficient exhibits a linear magnetic field-dependent behavior above $T_{\rm CDW}$.
		(b) Temperature dependence of Nernst coefficient. The Nernst coefficient shows an extreme value, which is marked by a dashed line. The temperature of the extreme value decreases as the magnetic field increases.
		(c) Temperature dependence of Hall resistivity divided by magnetic field. The compensation temperature of the Hall resistivity increases as the magnetic field increases.}
\end{figure*}

The electrical conductivity tensor $\overline{\sigma}$, thermal conductivity tensor $\overline{\kappa}$, and thermoelectric conductivity tensor $\overline{\alpha}$ relate the charge current $\textbf{J}_\textbf{e}$ and the heat current $\textbf{J}_\textbf{q}$ to the electric field $\textbf{E}$ and the thermal gradient $\nabla T$ \cite{Behnia2009}:
\begin{equation}\label{eq:AA}
\begin{aligned}
\left(
\begin{array}{c}
\mathbf{J}_\mathbf{e}\\
\mathbf{J}_\mathbf{q}
\end{array}
\right)
=
\left(
\begin{array}{cc}
\overline{\sigma} & - \overline{\alpha} \\
T \overline{\alpha} & - \overline{\kappa}
\end{array}
\right)
\left(
\begin{array}{c}
\mathbf{E}\\
\nabla T
\end{array}
\right)
\end{aligned}
\end{equation}
Electrons contribute to both electrical and thermal conductivity in metals, so $\overline{\sigma}$ relate to $\overline{\kappa}$ by the WF law, $\kappa_{\rm e} = L_0\sigma T$. $\kappa_{\rm e}$ is the thermal conductivity of electrons and $L_{0}$ is Lorentz constant, $L_{0} \equiv \frac{\pi^{2}}{3}\frac{k^{2}_{\rm B}}{e^{2}} = 2.44\times10^{-8} \ \rm{W \Omega K^{-2}}$.
$\overline{\sigma}$ can be measured by applying charge current without thermal gradient ($\nabla T = 0$). $\overline{\alpha}$ and $\overline{\kappa}$ can be measured by applying heat current without charge current ($\mathbf{J}_\mathbf{e} = 0$).
Using the configuration shown in Fig. \ref{fig:conductivity} (a), we measured the longitudinal thermal conductivity, the thermal Hall conductivity, the electrical hall conductivity, and the Nernst coefficient.

The temperature dependence of the longitudinal thermal conductivity $\kappa_{xx}$ is shown in Fig. \ref{fig:conductivity} (b). Above $T_{\rm CDW}$, $\kappa_{xx}$ decreases with decreasing temperature, which is different from ordinary materials, indicating that CDW fluctuations have a strong scattering effect on electron and phonon transport.
In the CDW order state, charge fluctuations are suppressed and the mean free path of electrons and phonons increases, so $\kappa_{xx}$ increases rapidly below $T_{\rm CDW}$ until it reaches a peak.
The total thermal conductivity is contributed by both phonons and electrons, $\kappa = \kappa_{\rm e} + \kappa_{\rm ph}$, where $\kappa_{\rm e}$ is the electron thermal conductivity and $\kappa_{\rm ph}$ is the phonon thermal conductivity \cite{Kim2012PhysRevB}.
In metals with short mean free path, the thermal conductivity of phonons should be of the same magnitude as that of electrons.
In order to compare the relationship between electrical conductivity and thermal conductivity, we converted electrical conductivity into electronic thermal conductivity, $L_0\sigma_{xx}T$, according to the WF law, as shown in Fig. \ref{fig:conductivity} (b). The total thermal conductivity is greater than $L_0\sigma_{xx}T$. The difference is due to the phonon contribution, which is of the same magnitude as the electron contribution, indicating that phonons play an important role in the thermal transport.
The thermal conductivity may also involve the influence of electron fluctuations on phonon scattering before and after CDW. The electron fluctuations near CDW phase transition will enhanced the phonon scattering, and the thermal conductivity of phonon is suppressed. So, the thermal conductivity increases slightly with the increase of temperature at $100-300$ K. As a result, WF law does not recover at $300$ K, and may recover at a higher temperature \cite{Kim2012PhysRevB, Kuo2006PhysRevB, Gumeniuk2015PhysRevB, Kuo2020PhysRevB, Kountz2021PhysRevB}.

The magnetic field dependence of longitudinal thermal conductivity $\kappa_{xx}$ and electrical conductivity $\sigma_{xx}$ at various temperatures are shown in Figs. \ref{fig:conductivity} (c) and \ref{fig:conductivity} (d) respectively. Above $40$ K, the field dependence of $\kappa_{xx}$ and and electrical conductivity $\sigma_{xx}$ are weak, while below $40$ K, the field dependence of $\kappa_{xx}$ and electrical conductivity $\sigma_{xx}$ become strong. The field dependence of longitudinal thermal conductivity is weaker than that of electronic conductivity, which also indicates the contribution of phonons to thermal transport.

Phonon thermal conductivity generally has no transverse effect, so the transverse thermal conductivity, namely thermal Hall conductivity, is a good method to study the thermal transport properties of electron.
Figures \ref{fig:Hall} (a) and \ref{fig:Hall} (b) show the field dependence of electrical Hall conductivity $\sigma_{xy}$ and thermal Hall conductivity $\kappa_{xy}$. Both $\sigma_{xy}$ and $\kappa_{xy}$ show linear field-dependent behaviors as conventional metals above $T_{\rm CDW}$.
Between $T_{\rm CDW}$ and $40$ K, $\kappa_{xy}$ is still linear magnetic field dependent in the high field region, but weakly nonlinear in the low field region. Below $40$ K, anomalous behaviors of thermal Hall conductivity arise. This anomalous behavior near 40 K is close related to the similar behaviors found in the muon spin relaxation rate, coherent phonon spectroscopy and Raman spectroscopy \cite{LiHaoxiang2021prx, Wang2021PhysRevB, Ratcliff2021PhysRevMaterials}, which have been recently proved to be an electronic nematic transition \cite{Nie2022nature}.
Below 30 K, there are two antisymmetric sideways “S” shape, as shown in Figs. \ref{fig:Hall} (a) and \ref{fig:Hall} (b). They exist in the high and low fields respectively. The antisymmetric sideways “S” shape is a characteristic of either an AHE or a two-band ordinary Hall effect. When the temperature rises above 30 K, the S-shaped hump in the high field disappears and changes linearly with the magnetic field, but the S-shaped hump in the low field still exists. It implies that the S-shape hump in the high field is the result of the two-band effect, and the S-shaped hump in the low field is related to ATHE \cite{Yang2020Hall, yu2021concurrence}.
We extract anomalous Hall conductivity $\sigma_{xy} ^{\rm AHE}$ and anomalous thermal Hall conductivity $\kappa_{xy} ^{\rm ATHE}$ by subtracting a local linear ordinary electrical Hall and thermal Hall background, as shown in Figs. \ref{fig:Hall} (d) and \ref{fig:Hall} (e).
The anomalous thermal Hall conductivity $\kappa_{xy} ^{\rm ATHE}$ increases rapidly with the magnetic field in the low field (below 1 T) and saturates above 1 T. The saturation value of $0.35 \ \rm{W K^{-1} m^{-1}}$ is in the range for metals with anomalous Hall effect and close to Co$_2$MnGa \cite{Xu2020PhysRevB, LiXiaokang2017PhysRevLett, ZhuZengwei2020}.
The anomalous thermal Hall ratio is comparable to that of anomalous Hall ratio (AHR) in $2$ T magnetic field, as shown in Fig. \ref{fig:Hall} (f), which is larger than the AHR of Fe, $\sim 0.8 \%$ \cite{Hou2015PRL}.
Anomalous Hall effect has been found in $A$V$_3$Sb$_5$ in previous studies on electric transport. By extracting $\sigma_{\rm{AHE}}$ at different angles of magnetic field relative to applied current, it is found that $\sigma_{\rm{AHE}}$ is not linearly proportional to the magnetic field component out of plane, which confirms the existence of AHE in $A$V$_3$Sb$_5$. For the kagome metal $A$V$_3$Sb$_5$ AHE, the current mainstream view is that AHE originates from the skew scattering effect \cite{Yang2020Hall, yu2021concurrence} and Berry curvature \cite{yu2021concurrence, gan2021magnetoseebeck}. The kagome sublattice in $A$V$_3$Sb$_5$ acts as a tilted spin cluster, which results in the enhancement of the skew scattering effect. In addition, $A$V$_3$Sb$_5$ has a large Berry curvature due to kagome lattice and its topological properties, which may be related to AHE.

Figure \ref{fig:Hall} (c) shows the temperature dependence of $\kappa_{xy}$ at $9$ T.
The thermal Hall effect is relatively weak at high temperature and becomes obvious with cooling.
As the temperature decreases, $\kappa_{xy}$ increases first and reaches a maximum around $40$ K, which happens to be the temperature at which anomalous behaviors begin to appear. As the temperature decreases further, $\kappa_{xy}$ changes sign at $27$ K, and reaches another extreme value at $12.5$ K.
In order to compare the thermal Hall conductivity and the electrical Hall conductivity, we calculate the electronic thermal Hall conductivity contribution $L_{0}\sigma_{xy}T$ by using the WF law, as shown in Fig. \ref{fig:Hall} (c). The behavior of $\kappa_{xy}$ is similar to $L_{0}\sigma_{xy}T$, but the value is slightly larger than $L_{0}\sigma_{xy}T$.
Phonons, as neutral quasiparticles with no charge, should not directly produce thermal Hall effect. It implies that phonon-drag mechanism or electronicly neutral topological low energy excitation may exist in the thermal Hall effect of the system \cite{LONG1967PLA}.
Combined with the large phonon contribution in $\kappa_{xx}$, a strong electron-phonon coupling in CsV$_{3}$Sb$_{5}$ was observed from our thermal measurements. Since the superconductivity is one of the most important properties in the $A$V$_3$Sb$_5$ materials, what is the driving force for SC becomes one central issue. The electron-phonon coupling observed here further supports the superconductivity of CsV$_3$Sb$_5$ is a conventional $s$ wave \cite{Luo2022NC,duan2021nodeless,Mu2021CPL,xu2021multiband}.

In order to further study the anomalous transverse effect, we carried out the Nernst effect measurement.
Equation (\ref{eq:AA}) yields
\begin{equation}\label{eq:CC}
	\mathbf{E} = {\bar{\sigma}}^{-1} \bar{\alpha}  \nabla T
\end{equation}
For a single band metal, the Nernst coefficient derived from Eq. (\ref{eq:CC}) is
\begin{equation}\label{eq:DD}
N = \frac{E_{y}}{\nabla _x T} = \frac{\alpha_{xy}\sigma_{xx} - \alpha_{xx}\sigma_{xy}}{\sigma_{xx}^{2} + \sigma_{xy}^{2}}
\end{equation}
If the electric conductivity does not depend on energy, then
\begin{equation}\label{eq:EE}
\frac{\sigma_{xy}}{\sigma_{xx}} = \frac{\alpha_{xy}}{\alpha_{xx}}
\end{equation}
Therefore, the two terms in Eq. (\ref{eq:DD}) are eliminated, which is called ``Sondheimer cancellation.''
As shown in Fig. \ref{fig:schematic} (a),
the charge current generated by the thermal gradient will accumulate charges and create an electric field. The charge current produced by the thermal gradient and electric field offsets each other.
However, this cancellation does not affect the electrical Hall effect.
On the other hand, ambipolar flow can lead to an enhancement of the Nernst signal in multiband metals \cite{Behnia2009,NbSe22003}. If we assume two types of carriers, Eq. (\ref{eq:DD}) is rewritten as
\begin{equation}\label{eq:FF}
N = \frac{(\alpha_{xy}^+ + \alpha_{xy}^-)(\sigma_{xx}^+ + \sigma_{xx}^-) - (\alpha_{xx}^+ + \alpha_{xx}^-)(\sigma_{xy}^+ + \sigma_{xy}^-)}{(\sigma_{xx}^+ + \sigma_{xx}^-)^{2}+(\sigma_{xy}^+ + \sigma_{xy}^-)^{2}}
\end{equation}
The superscripts $+$ and $-$ indicate the hole-like and the electron-like carriers. The $\sigma_{xy}$ of two type carries have opposite signs, so the Sondheimer cancellation is avoid.
Especially when $\sigma_{xy}^- = -\sigma_{xy}^+$, Eq. (\ref{eq:FF}) has a maximum, which means that the Nernst coefficient will show a maximum when the Hall effect is suppressed, as shown in Fig. \ref{fig:schematic}.

The field dependence of Nernst coefficient $N$ in CsV$_3$Sb$_5$ at various temperatures is shown in Fig. \ref{fig:Nernst} (a).
Similar to Hall resistivity, the Nernst coefficient exhibits linear field-dependent behavior above $T_{\rm CDW}$. When the temperature is lower than $T_{\rm CDW}$, the Nernst coefficient gradually deviates from the linear magnetic field dependence and the anomalous behavior show up below $40$ K.
The temperature variation of Nenst coefficient and Hall resistance in various magnetic fields are shown in Figs. \ref{fig:Nernst} (b) and \ref{fig:Nernst} (c) respectively.
At $9$ T, the Nenst coefficient reaches its maximum value at $32$ K where $\rho_{xy}=0$, which indicates a multiband effect. When the magnetic field decreases, the temperature of the extreme value of Nernst coefficient increases and reaches $40$ K at zero field limit, while the compensation temperature of Hall resistivity decreases and approaches zero at zero field limit. Hence, there are other factors that influence the transport properties.
Nernst and Hall signals are dominated by normal linear items in high magnetic fields and dominated by anomalous items in low magnetic fields.
Therefore, the multiband model should not be the only reason for the relationship between Hall resistivity and Nernst coefficient in low magnetic fields. One possible reason is that a new small electronic pocket with high mobility appears under CDW \cite{Zhu2011JOP, Behnia2009}. But there is no direct experimental evidence for the new small electronic pocket with high mobility after CDW phase transition in $A$V$_3$Sb$_5$, which can be seen from the high-resolution ARPES measurements. \cite{Ortiz2020Cs, Congli2021spectroscopic, Luo2022NC, Nakayama2021PRB, Lou2022PRL, Liu2021PRX}.
In the anomalous Hall effect studies of $A$V$_3$Sb$_5$ \cite{Yang2020Hall, yu2021concurrence}, the skew scattering has been widely proposed. However, since the skew scattering does not break Sondheimer cancellation, the finite Nernst signal in low fields should be dominated by other origins, like the Berry curvature from the $Z_2$ topological property and time-reversal symmetry breaking in CsV$_3$Sb$_5$ \cite{XiaoDi2006PHYSICALREVIEWLETTERS,McCormick2017prb}.
The expressions of anomalous Hall conductivity $\sigma_{ij}$ and anomalous thermal Hall conductivity $\kappa_{ij}$ can be obtained according to theoretical derivation \cite{Wang2021arxiv, ZhuZengwei2020}:
\begin{equation}\label{eq:GG}
\sigma_{ij}^A(\mu) = \frac{e^2}{\hbar}\int_{-\infty}^{\infty}d\xi(-\frac{\partial f(\xi - \mu)}{\partial \xi})\widetilde{\sigma}_{ij}(\xi)
\end{equation}
\begin{equation}\label{eq:HH}
	\kappa_{ij}^A(\mu) = \frac{1}{\hbar T}\int_{-\infty}^{\infty}d\xi(-(\xi - \mu)^2\frac{\partial f(\xi - \mu)}{\partial \xi})\widetilde{\sigma}_{ij}(\xi)
\end{equation}
\begin{equation}\label{eq:II}
\widetilde{\sigma}_{ij}(\xi) = \int_{\rm BZ}\frac{d \textbf{k}}{(2 \pi)^3}\sum_{\epsilon_n \textless \xi}\Omega_{ij}^n(\textbf{k})
\end{equation}
where $\Omega_{ij}^n(\textbf{k})$ is Berry curvature and $f(\xi - \mu) = 1/(e^(\frac{\xi - \mu}{k_{\rm{B}}T})+1)$ is the Fermi-Dirac function. From the comparison of Eqs. (\ref{eq:GG}) and (\ref{eq:HH}), we can find that the Berry curvature contributions to the thermal and electrical transport have different pondering functions, which leads to the mismatch between $\sigma_{ij}$ and $\kappa_{ij}$ \cite{ZhuZengwei2020, Linchao2021JPS}. These differences reflect the fact that Berry curvatures from different regions of reciprocal space carry opposite signs and thus compete, which may cause the WF ratio to increase or decrease \cite{Wang2021arxiv}.

\section{Summary}
In summary, we performed the thermal transport measurements on CsV$_3$Sb$_5$. Both the longitudinal thermal conductivity and the thermal Hall conductivity are larger than the theoretical values, indicating that there is a large phonon contribution in thermal transports. This shows that the phonon plays an important role in the physics of  CsV$_3$Sb$_5$ even at low temperature, which is consistent with the electron-phonon coupling  found in ARPES \cite{Luo2022NC} and a conventional $s$-wave superconductor from NMR \cite{Mu2021CPL} and penetration depth measurement \cite{duan2021nodeless}.
The anomalous thermal Hall effect and the anomalous Nernst effect appear in the CDW state.
The Nernst coefficient reaches a maximum value when the Hall resistivity is zero at $9$ T, which indicates a multiband effect domination. As the magnetic field decreases, the maximum of Nernst effect shifts to higher temperature, while the zero value of Hall resistivity shifts to lower temperature. It indicates that the anomalous transverse response in the low field may have a contribution from the Berry curvature.

\emph{Note Added}. During the preparation of this manuscript, we realized that a similar anomalous Nernst effect on CsV$_3$Sb$_5$ has been posted in Ref. \cite{gan2021magnetoseebeck}, which is consistent with our results.

\begin{acknowledgments}
This work was supported by the National Key Research and Development Program of China (Grant No. 2017YFA0302901), the National Science Foundation of China (Grants No. 12134018, No. 11921004, No. 11634015), the Strategic Priority Research Program and Key Research Program of Frontier Sciences of the Chinese Academy of Sciences (Grant No. XDB33010100), and the Synergetic Extreme Condition User Facility (SECUF).
\end{acknowledgments}




%

\begin{thebibliography}{57}%
\makeatletter
\providecommand \@ifxundefined [1]{%
 \@ifx{#1\undefined}
}%
\providecommand \@ifnum [1]{%
 \ifnum #1\expandafter \@firstoftwo
 \else \expandafter \@secondoftwo
 \fi
}%
\providecommand \@ifx [1]{%
 \ifx #1\expandafter \@firstoftwo
 \else \expandafter \@secondoftwo
 \fi
}%
\providecommand \natexlab [1]{#1}%
\providecommand \enquote  [1]{``#1''}%
\providecommand \bibnamefont  [1]{#1}%
\providecommand \bibfnamefont [1]{#1}%
\providecommand \citenamefont [1]{#1}%
\providecommand \href@noop [0]{\@secondoftwo}%
\providecommand \href [0]{\begingroup \@sanitize@url \@href}%
\providecommand \@href[1]{\@@startlink{#1}\@@href}%
\providecommand \@@href[1]{\endgroup#1\@@endlink}%
\providecommand \@sanitize@url [0]{\catcode `\\12\catcode `\$12\catcode
  `\&12\catcode `\#12\catcode `\^12\catcode `\_12\catcode `\%12\relax}%
\providecommand \@@startlink[1]{}%
\providecommand \@@endlink[0]{}%
\providecommand \url  [0]{\begingroup\@sanitize@url \@url }%
\providecommand \@url [1]{\endgroup\@href {#1}{\urlprefix }}%
\providecommand \urlprefix  [0]{URL }%
\providecommand \Eprint [0]{\href }%
\providecommand \doibase [0]{https://doi.org/}%
\providecommand \selectlanguage [0]{\@gobble}%
\providecommand \bibinfo  [0]{\@secondoftwo}%
\providecommand \bibfield  [0]{\@secondoftwo}%
\providecommand \translation [1]{[#1]}%
\providecommand \BibitemOpen [0]{}%
\providecommand \bibitemStop [0]{}%
\providecommand \bibitemNoStop [0]{.\EOS\space}%
\providecommand \EOS [0]{\spacefactor3000\relax}%
\providecommand \BibitemShut  [1]{\csname bibitem#1\endcsname}%
\let\auto@bib@innerbib\@empty
\bibitem [{\citenamefont {Ortiz}\ \emph {et~al.}(2019)\citenamefont {Ortiz},
  \citenamefont {Gomes}, \citenamefont {Morey}, \citenamefont {Winiarski},
  \citenamefont {Bordelon}, \citenamefont {Mangum}, \citenamefont {Oswald},
  \citenamefont {Rodriguez-Rivera}, \citenamefont {Neilson}, \citenamefont
  {Wilson}, \citenamefont {Ertekin}, \citenamefont {McQueen},\ and\
  \citenamefont {Toberer}}]{Ortiz2019New}%
  \BibitemOpen
  \bibfield  {author} {\bibinfo {author} {\bibfnamefont {B.~R.}\ \bibnamefont
  {Ortiz}}, \bibinfo {author} {\bibfnamefont {L.~C.}\ \bibnamefont {Gomes}},
  \bibinfo {author} {\bibfnamefont {J.~R.}\ \bibnamefont {Morey}}, \bibinfo
  {author} {\bibfnamefont {M.}~\bibnamefont {Winiarski}}, \bibinfo {author}
  {\bibfnamefont {M.}~\bibnamefont {Bordelon}}, \bibinfo {author}
  {\bibfnamefont {J.~S.}\ \bibnamefont {Mangum}}, \bibinfo {author}
  {\bibfnamefont {I.~W.~H.}\ \bibnamefont {Oswald}}, \bibinfo {author}
  {\bibfnamefont {J.~A.}\ \bibnamefont {Rodriguez-Rivera}}, \bibinfo {author}
  {\bibfnamefont {J.~R.}\ \bibnamefont {Neilson}}, \bibinfo {author}
  {\bibfnamefont {S.~D.}\ \bibnamefont {Wilson}}, \bibinfo {author}
  {\bibfnamefont {E.}~\bibnamefont {Ertekin}}, \bibinfo {author} {\bibfnamefont
  {T.~M.}\ \bibnamefont {McQueen}},\ and\ \bibinfo {author} {\bibfnamefont
  {E.~S.}\ \bibnamefont {Toberer}},\ }\bibfield  {title} {\bibinfo {title}
  {{New kagome prototype materials: discovery of KV$_{3}$Sb$_{5}$,
  RbV$_{3}$Sb$_{5}$, and CsV$_{3}$Sb$_{5}$}},\ }\href
  {https://doi.org/10.1103/PhysRevMaterials.3.094407} {\bibfield  {journal}
  {\bibinfo  {journal} {Phys. Rev. Materials}\ }\textbf {\bibinfo {volume}
  {3}},\ \bibinfo {pages} {094407} (\bibinfo {year} {2019})}\BibitemShut
  {NoStop}%
\bibitem [{\citenamefont {Ortiz}\ \emph {et~al.}(2020)\citenamefont {Ortiz},
  \citenamefont {Teicher}, \citenamefont {Hu}, \citenamefont {Zuo},
  \citenamefont {Sarte}, \citenamefont {Schueller}, \citenamefont {Abeykoon},
  \citenamefont {Krogstad}, \citenamefont {Rosenkranz}, \citenamefont {Osborn},
  \citenamefont {Seshadri}, \citenamefont {Balents}, \citenamefont {He},\ and\
  \citenamefont {Wilson}}]{Ortiz2020Cs}%
  \BibitemOpen
  \bibfield  {author} {\bibinfo {author} {\bibfnamefont {B.~R.}\ \bibnamefont
  {Ortiz}}, \bibinfo {author} {\bibfnamefont {S.~M.~L.}\ \bibnamefont
  {Teicher}}, \bibinfo {author} {\bibfnamefont {Y.}~\bibnamefont {Hu}},
  \bibinfo {author} {\bibfnamefont {J.~L.}\ \bibnamefont {Zuo}}, \bibinfo
  {author} {\bibfnamefont {P.~M.}\ \bibnamefont {Sarte}}, \bibinfo {author}
  {\bibfnamefont {E.~C.}\ \bibnamefont {Schueller}}, \bibinfo {author}
  {\bibfnamefont {A.~M.~M.}\ \bibnamefont {Abeykoon}}, \bibinfo {author}
  {\bibfnamefont {M.~J.}\ \bibnamefont {Krogstad}}, \bibinfo {author}
  {\bibfnamefont {S.}~\bibnamefont {Rosenkranz}}, \bibinfo {author}
  {\bibfnamefont {R.}~\bibnamefont {Osborn}}, \bibinfo {author} {\bibfnamefont
  {R.}~\bibnamefont {Seshadri}}, \bibinfo {author} {\bibfnamefont
  {L.}~\bibnamefont {Balents}}, \bibinfo {author} {\bibfnamefont
  {J.}~\bibnamefont {He}},\ and\ \bibinfo {author} {\bibfnamefont {S.~D.}\
  \bibnamefont {Wilson}},\ }\bibfield  {title} {\bibinfo {title}
  {{CsV$_{3}$Sb$_{5}$: A Z$_{2}$ topological kagome metal with a
  superconducting ground state}},\ }\href
  {https://doi.org/10.1103/PhysRevLett.125.247002} {\bibfield  {journal}
  {\bibinfo  {journal} {Phys. Rev. Lett.}\ }\textbf {\bibinfo {volume} {125}},\
  \bibinfo {pages} {247002} (\bibinfo {year} {2020})}\BibitemShut {NoStop}%
\bibitem [{\citenamefont {Ortiz}\ \emph {et~al.}(2021)\citenamefont {Ortiz},
  \citenamefont {Sarte}, \citenamefont {Kenney}, \citenamefont {Graf},
  \citenamefont {Teicher}, \citenamefont {Seshadri},\ and\ \citenamefont
  {Wilson}}]{Ortiz2021KVSb}%
  \BibitemOpen
  \bibfield  {author} {\bibinfo {author} {\bibfnamefont {B.~R.}\ \bibnamefont
  {Ortiz}}, \bibinfo {author} {\bibfnamefont {P.~M.}\ \bibnamefont {Sarte}},
  \bibinfo {author} {\bibfnamefont {E.~M.}\ \bibnamefont {Kenney}}, \bibinfo
  {author} {\bibfnamefont {M.~J.}\ \bibnamefont {Graf}}, \bibinfo {author}
  {\bibfnamefont {S.~M.~L.}\ \bibnamefont {Teicher}}, \bibinfo {author}
  {\bibfnamefont {R.}~\bibnamefont {Seshadri}},\ and\ \bibinfo {author}
  {\bibfnamefont {S.~D.}\ \bibnamefont {Wilson}},\ }\bibfield  {title}
  {\bibinfo {title} {{Superconductivity in the ${\mathbb{Z}}_{2}$ kagome metal
  KV$_{3}$Sb$_{5}$}},\ }\href
  {https://doi.org/10.1103/PhysRevMaterials.5.034801} {\bibfield  {journal}
  {\bibinfo  {journal} {Phys. Rev. Materials}\ }\textbf {\bibinfo {volume}
  {5}},\ \bibinfo {pages} {034801} (\bibinfo {year} {2021})}\BibitemShut
  {NoStop}%
\bibitem [{\citenamefont {Yin}\ \emph {et~al.}(2021)\citenamefont {Yin},
  \citenamefont {Tu}, \citenamefont {Gong}, \citenamefont {Fu}, \citenamefont
  {Yan},\ and\ \citenamefont {Lei}}]{Lei2021Rb}%
  \BibitemOpen
  \bibfield  {author} {\bibinfo {author} {\bibfnamefont {Q.~W.}\ \bibnamefont
  {Yin}}, \bibinfo {author} {\bibfnamefont {Z.~J.}\ \bibnamefont {Tu}},
  \bibinfo {author} {\bibfnamefont {C.~S.}\ \bibnamefont {Gong}}, \bibinfo
  {author} {\bibfnamefont {Y.}~\bibnamefont {Fu}}, \bibinfo {author}
  {\bibfnamefont {S.~H.}\ \bibnamefont {Yan}},\ and\ \bibinfo {author}
  {\bibfnamefont {H.~C.}\ \bibnamefont {Lei}},\ }\bibfield  {title} {\bibinfo
  {title} {{Superconductivity and Normal-State Properties of Kagome Metal
  RbV$_{3}$Sb$_{5}$ Single Crystals}},\ }\href
  {https://doi.org/10.1088/0256-307x/38/3/037403} {\bibfield  {journal}
  {\bibinfo  {journal} {Chinese Physics Letters}\ }\textbf {\bibinfo {volume}
  {38}},\ \bibinfo {pages} {6} (\bibinfo {year} {2021})}\BibitemShut {NoStop}%
\bibitem [{\citenamefont {Jiang}\ \emph
  {et~al.}(2021{\natexlab{a}})\citenamefont {Jiang}, \citenamefont {Wu},
  \citenamefont {Yin}, \citenamefont {Wang}, \citenamefont {Hasan},
  \citenamefont {Wilson}, \citenamefont {Chen},\ and\ \citenamefont
  {Hu}}]{jiang2021kagome}%
  \BibitemOpen
  \bibfield  {author} {\bibinfo {author} {\bibfnamefont {K.}~\bibnamefont
  {Jiang}}, \bibinfo {author} {\bibfnamefont {T.}~\bibnamefont {Wu}}, \bibinfo
  {author} {\bibfnamefont {J.-X.}\ \bibnamefont {Yin}}, \bibinfo {author}
  {\bibfnamefont {Z.}~\bibnamefont {Wang}}, \bibinfo {author} {\bibfnamefont
  {M.~Z.}\ \bibnamefont {Hasan}}, \bibinfo {author} {\bibfnamefont {S.~D.}\
  \bibnamefont {Wilson}}, \bibinfo {author} {\bibfnamefont {X.}~\bibnamefont
  {Chen}},\ and\ \bibinfo {author} {\bibfnamefont {J.}~\bibnamefont {Hu}},\
  }\href@noop {} {\bibinfo {title} {{Kagome superconductors AV$_3$Sb$_5$ (A=K,
  Rb, Cs)}}} (\bibinfo {year} {2021}{\natexlab{a}}),\ \Eprint
  {https://arxiv.org/abs/2109.10809} {arXiv:2109.10809 [cond-mat.supr-con]}
  \BibitemShut {NoStop}%
\bibitem [{\citenamefont {Zhao}\ \emph
  {et~al.}(2021{\natexlab{a}})\citenamefont {Zhao}, \citenamefont {Li},
  \citenamefont {Ortiz}, \citenamefont {Teicher}, \citenamefont {Park},
  \citenamefont {Ye}, \citenamefont {Wang}, \citenamefont {Balents},
  \citenamefont {Wilson},\ and\ \citenamefont {Zeljkovic}}]{zhao2021cascade}%
  \BibitemOpen
  \bibfield  {author} {\bibinfo {author} {\bibfnamefont {H.}~\bibnamefont
  {Zhao}}, \bibinfo {author} {\bibfnamefont {H.}~\bibnamefont {Li}}, \bibinfo
  {author} {\bibfnamefont {B.~R.}\ \bibnamefont {Ortiz}}, \bibinfo {author}
  {\bibfnamefont {S.~M.~L.}\ \bibnamefont {Teicher}}, \bibinfo {author}
  {\bibfnamefont {T.}~\bibnamefont {Park}}, \bibinfo {author} {\bibfnamefont
  {M.}~\bibnamefont {Ye}}, \bibinfo {author} {\bibfnamefont {Z.}~\bibnamefont
  {Wang}}, \bibinfo {author} {\bibfnamefont {L.}~\bibnamefont {Balents}},
  \bibinfo {author} {\bibfnamefont {S.~D.}\ \bibnamefont {Wilson}},\ and\
  \bibinfo {author} {\bibfnamefont {I.}~\bibnamefont {Zeljkovic}},\ }\href@noop
  {} {\bibinfo {title} {{Cascade of correlated electron states in a kagome
  superconductor CsV$_{3}$Sb$_{5}$}}} (\bibinfo {year} {2021}{\natexlab{a}}),\
  \Eprint {https://arxiv.org/abs/2103.03118} {arXiv:2103.03118
  [cond-mat.supr-con]} \BibitemShut {NoStop}%
\bibitem [{\citenamefont {Zhao}\ \emph
  {et~al.}(2021{\natexlab{b}})\citenamefont {Zhao}, \citenamefont {Wang},
  \citenamefont {Xia}, \citenamefont {Yin}, \citenamefont {Ni}, \citenamefont
  {Huang}, \citenamefont {Tu}, \citenamefont {Tao}, \citenamefont {Tu},
  \citenamefont {Gong}, \citenamefont {Lei}, \citenamefont {Guo}, \citenamefont
  {Yang},\ and\ \citenamefont {Li}}]{zhao2021nodal}%
  \BibitemOpen
  \bibfield  {author} {\bibinfo {author} {\bibfnamefont {C.~C.}\ \bibnamefont
  {Zhao}}, \bibinfo {author} {\bibfnamefont {L.~S.}\ \bibnamefont {Wang}},
  \bibinfo {author} {\bibfnamefont {W.}~\bibnamefont {Xia}}, \bibinfo {author}
  {\bibfnamefont {Q.~W.}\ \bibnamefont {Yin}}, \bibinfo {author} {\bibfnamefont
  {J.~M.}\ \bibnamefont {Ni}}, \bibinfo {author} {\bibfnamefont {Y.~Y.}\
  \bibnamefont {Huang}}, \bibinfo {author} {\bibfnamefont {C.~P.}\ \bibnamefont
  {Tu}}, \bibinfo {author} {\bibfnamefont {Z.~C.}\ \bibnamefont {Tao}},
  \bibinfo {author} {\bibfnamefont {Z.~J.}\ \bibnamefont {Tu}}, \bibinfo
  {author} {\bibfnamefont {C.~S.}\ \bibnamefont {Gong}}, \bibinfo {author}
  {\bibfnamefont {H.~C.}\ \bibnamefont {Lei}}, \bibinfo {author} {\bibfnamefont
  {Y.~F.}\ \bibnamefont {Guo}}, \bibinfo {author} {\bibfnamefont {X.~F.}\
  \bibnamefont {Yang}},\ and\ \bibinfo {author} {\bibfnamefont {S.~Y.}\
  \bibnamefont {Li}},\ }\href@noop {} {\bibinfo {title} {{Nodal
  superconductivity and superconducting domes in the topological Kagome metal
  CsV$_{3}$Sb$_{5}$}}} (\bibinfo {year} {2021}{\natexlab{b}}),\ \Eprint
  {https://arxiv.org/abs/2102.08356} {arXiv:2102.08356 [cond-mat.supr-con]}
  \BibitemShut {NoStop}%
\bibitem [{\citenamefont {Chen}\ \emph {et~al.}(2021)\citenamefont {Chen},
  \citenamefont {Yang}, \citenamefont {Hu}, \citenamefont {Zhao}, \citenamefont
  {Yuan}, \citenamefont {Xing}, \citenamefont {Qian}, \citenamefont {Huang},
  \citenamefont {Li}, \citenamefont {Ye}, \citenamefont {Ma}, \citenamefont
  {Ni}, \citenamefont {Zhang}, \citenamefont {Yin}, \citenamefont {Gong},
  \citenamefont {Tu}, \citenamefont {Lei}, \citenamefont {Tan}, \citenamefont
  {Zhou}, \citenamefont {Shen}, \citenamefont {Dong}, \citenamefont {Yan},
  \citenamefont {Wang},\ and\ \citenamefont {Gao}}]{chen2021roton}%
  \BibitemOpen
  \bibfield  {author} {\bibinfo {author} {\bibfnamefont {H.}~\bibnamefont
  {Chen}}, \bibinfo {author} {\bibfnamefont {H.}~\bibnamefont {Yang}}, \bibinfo
  {author} {\bibfnamefont {B.}~\bibnamefont {Hu}}, \bibinfo {author}
  {\bibfnamefont {Z.}~\bibnamefont {Zhao}}, \bibinfo {author} {\bibfnamefont
  {J.}~\bibnamefont {Yuan}}, \bibinfo {author} {\bibfnamefont {Y.}~\bibnamefont
  {Xing}}, \bibinfo {author} {\bibfnamefont {G.}~\bibnamefont {Qian}}, \bibinfo
  {author} {\bibfnamefont {Z.}~\bibnamefont {Huang}}, \bibinfo {author}
  {\bibfnamefont {G.}~\bibnamefont {Li}}, \bibinfo {author} {\bibfnamefont
  {Y.}~\bibnamefont {Ye}}, \bibinfo {author} {\bibfnamefont {S.}~\bibnamefont
  {Ma}}, \bibinfo {author} {\bibfnamefont {S.}~\bibnamefont {Ni}}, \bibinfo
  {author} {\bibfnamefont {H.}~\bibnamefont {Zhang}}, \bibinfo {author}
  {\bibfnamefont {Q.}~\bibnamefont {Yin}}, \bibinfo {author} {\bibfnamefont
  {C.}~\bibnamefont {Gong}}, \bibinfo {author} {\bibfnamefont {Z.}~\bibnamefont
  {Tu}}, \bibinfo {author} {\bibfnamefont {H.}~\bibnamefont {Lei}}, \bibinfo
  {author} {\bibfnamefont {H.}~\bibnamefont {Tan}}, \bibinfo {author}
  {\bibfnamefont {S.}~\bibnamefont {Zhou}}, \bibinfo {author} {\bibfnamefont
  {C.}~\bibnamefont {Shen}}, \bibinfo {author} {\bibfnamefont {X.}~\bibnamefont
  {Dong}}, \bibinfo {author} {\bibfnamefont {B.}~\bibnamefont {Yan}}, \bibinfo
  {author} {\bibfnamefont {Z.}~\bibnamefont {Wang}},\ and\ \bibinfo {author}
  {\bibfnamefont {H.-J.}\ \bibnamefont {Gao}},\ }\bibfield  {title} {\bibinfo
  {title} {{Roton pair density wave in a strong-coupling kagome
  superconductor}},\ }\href {https://doi.org/10.1038/s41586-021-03983-5}
  {\bibfield  {journal} {\bibinfo  {journal} {Nature}\ }\textbf {\bibinfo
  {volume} {599}},\ \bibinfo {pages} {222} (\bibinfo {year}
  {2021})}\BibitemShut {NoStop}%
\bibitem [{\citenamefont {Mu}\ \emph {et~al.}(2021)\citenamefont {Mu},
  \citenamefont {Yin}, \citenamefont {Tu}, \citenamefont {Gong}, \citenamefont
  {Lei}, \citenamefont {Li},\ and\ \citenamefont {Luo}}]{Mu2021CPL}%
  \BibitemOpen
  \bibfield  {author} {\bibinfo {author} {\bibfnamefont {C.}~\bibnamefont
  {Mu}}, \bibinfo {author} {\bibfnamefont {Q.}~\bibnamefont {Yin}}, \bibinfo
  {author} {\bibfnamefont {Z.}~\bibnamefont {Tu}}, \bibinfo {author}
  {\bibfnamefont {C.}~\bibnamefont {Gong}}, \bibinfo {author} {\bibfnamefont
  {H.}~\bibnamefont {Lei}}, \bibinfo {author} {\bibfnamefont {Z.}~\bibnamefont
  {Li}},\ and\ \bibinfo {author} {\bibfnamefont {J.}~\bibnamefont {Luo}},\
  }\bibfield  {title} {\bibinfo {title} {{S-Wave Superconductivity in Kagome
  Metal CsV$_{3}$Sb$_{5}$ Revealed by $^{121/123}$Sb NQR and $^{51}$V NMR
  Measurements}},\ }\href {https://doi.org/10.1088/0256-307X/38/7/077402}
  {\bibfield  {journal} {\bibinfo  {journal} {Chinese Physics Letters}\
  }\textbf {\bibinfo {volume} {38}},\ \bibinfo {eid} {077402} (\bibinfo {year}
  {2021})}\BibitemShut {NoStop}%
\bibitem [{\citenamefont {Duan}\ \emph {et~al.}(2021)\citenamefont {Duan},
  \citenamefont {Nie}, \citenamefont {Luo}, \citenamefont {Yu}, \citenamefont
  {Ortiz}, \citenamefont {Yin}, \citenamefont {Su}, \citenamefont {Du},
  \citenamefont {Wang}, \citenamefont {Chen}, \citenamefont {Lu}, \citenamefont
  {Ying}, \citenamefont {Wilson}, \citenamefont {Chen}, \citenamefont {Song},\
  and\ \citenamefont {Yuan}}]{duan2021nodeless}%
  \BibitemOpen
  \bibfield  {author} {\bibinfo {author} {\bibfnamefont {W.}~\bibnamefont
  {Duan}}, \bibinfo {author} {\bibfnamefont {Z.}~\bibnamefont {Nie}}, \bibinfo
  {author} {\bibfnamefont {S.}~\bibnamefont {Luo}}, \bibinfo {author}
  {\bibfnamefont {F.}~\bibnamefont {Yu}}, \bibinfo {author} {\bibfnamefont
  {B.~R.}\ \bibnamefont {Ortiz}}, \bibinfo {author} {\bibfnamefont
  {L.}~\bibnamefont {Yin}}, \bibinfo {author} {\bibfnamefont {H.}~\bibnamefont
  {Su}}, \bibinfo {author} {\bibfnamefont {F.}~\bibnamefont {Du}}, \bibinfo
  {author} {\bibfnamefont {A.}~\bibnamefont {Wang}}, \bibinfo {author}
  {\bibfnamefont {Y.}~\bibnamefont {Chen}}, \bibinfo {author} {\bibfnamefont
  {X.}~\bibnamefont {Lu}}, \bibinfo {author} {\bibfnamefont {J.}~\bibnamefont
  {Ying}}, \bibinfo {author} {\bibfnamefont {S.~D.}\ \bibnamefont {Wilson}},
  \bibinfo {author} {\bibfnamefont {X.}~\bibnamefont {Chen}}, \bibinfo {author}
  {\bibfnamefont {Y.}~\bibnamefont {Song}},\ and\ \bibinfo {author}
  {\bibfnamefont {H.}~\bibnamefont {Yuan}},\ }\bibfield  {title} {\bibinfo
  {title} {{Nodeless superconductivity in the kagome metal
  CsV$_{3}$Sb$_{5}$}},\ }\href {https://doi.org/10.1007/s11433-021-1747-7}
  {\bibfield  {journal} {\bibinfo  {journal} {{Science China Physics, Mechanics
  $\&$ Astronomy}}\ }\textbf {\bibinfo {volume} {64}},\ \bibinfo {pages}
  {107462} (\bibinfo {year} {2021})}\BibitemShut {NoStop}%
\bibitem [{\citenamefont {Xu}\ \emph {et~al.}(2021)\citenamefont {Xu},
  \citenamefont {Yan}, \citenamefont {Yin}, \citenamefont {Xia}, \citenamefont
  {Fang}, \citenamefont {Chen}, \citenamefont {Li}, \citenamefont {Yang},
  \citenamefont {Guo},\ and\ \citenamefont {Feng}}]{xu2021multiband}%
  \BibitemOpen
  \bibfield  {author} {\bibinfo {author} {\bibfnamefont {H.-S.}\ \bibnamefont
  {Xu}}, \bibinfo {author} {\bibfnamefont {Y.-J.}\ \bibnamefont {Yan}},
  \bibinfo {author} {\bibfnamefont {R.}~\bibnamefont {Yin}}, \bibinfo {author}
  {\bibfnamefont {W.}~\bibnamefont {Xia}}, \bibinfo {author} {\bibfnamefont
  {S.}~\bibnamefont {Fang}}, \bibinfo {author} {\bibfnamefont {Z.}~\bibnamefont
  {Chen}}, \bibinfo {author} {\bibfnamefont {Y.}~\bibnamefont {Li}}, \bibinfo
  {author} {\bibfnamefont {W.}~\bibnamefont {Yang}}, \bibinfo {author}
  {\bibfnamefont {Y.}~\bibnamefont {Guo}},\ and\ \bibinfo {author}
  {\bibfnamefont {D.-L.}\ \bibnamefont {Feng}},\ }\bibfield  {title} {\bibinfo
  {title} {{Multiband Superconductivity with Sign-Preserving Order Parameter in
  Kagome Superconductor ${\mathrm{CsV}}_{3}{\mathrm{Sb}}_{5}$}},\ }\href
  {https://doi.org/10.1103/PhysRevLett.127.187004} {\bibfield  {journal}
  {\bibinfo  {journal} {Phys. Rev. Lett.}\ }\textbf {\bibinfo {volume} {127}},\
  \bibinfo {pages} {187004} (\bibinfo {year} {2021})}\BibitemShut {NoStop}%
\bibitem [{\citenamefont {Luo}\ \emph {et~al.}(2022)\citenamefont {Luo},
  \citenamefont {Gao}, \citenamefont {Liu}, \citenamefont {Gu}, \citenamefont
  {Wu}, \citenamefont {Yi}, \citenamefont {Jia}, \citenamefont {Wu},
  \citenamefont {Luo}, \citenamefont {Xu}, \citenamefont {Zhao}, \citenamefont
  {Wang}, \citenamefont {Mao}, \citenamefont {Liu}, \citenamefont {Zhu},
  \citenamefont {Shi}, \citenamefont {Jiang}, \citenamefont {Hu}, \citenamefont
  {Xu},\ and\ \citenamefont {Zhou}}]{Luo2022NC}%
  \BibitemOpen
  \bibfield  {author} {\bibinfo {author} {\bibfnamefont {H.}~\bibnamefont
  {Luo}}, \bibinfo {author} {\bibfnamefont {Q.}~\bibnamefont {Gao}}, \bibinfo
  {author} {\bibfnamefont {H.}~\bibnamefont {Liu}}, \bibinfo {author}
  {\bibfnamefont {Y.}~\bibnamefont {Gu}}, \bibinfo {author} {\bibfnamefont
  {D.}~\bibnamefont {Wu}}, \bibinfo {author} {\bibfnamefont {C.}~\bibnamefont
  {Yi}}, \bibinfo {author} {\bibfnamefont {J.}~\bibnamefont {Jia}}, \bibinfo
  {author} {\bibfnamefont {S.}~\bibnamefont {Wu}}, \bibinfo {author}
  {\bibfnamefont {X.}~\bibnamefont {Luo}}, \bibinfo {author} {\bibfnamefont
  {Y.}~\bibnamefont {Xu}}, \bibinfo {author} {\bibfnamefont {L.}~\bibnamefont
  {Zhao}}, \bibinfo {author} {\bibfnamefont {Q.}~\bibnamefont {Wang}}, \bibinfo
  {author} {\bibfnamefont {H.}~\bibnamefont {Mao}}, \bibinfo {author}
  {\bibfnamefont {G.}~\bibnamefont {Liu}}, \bibinfo {author} {\bibfnamefont
  {Z.}~\bibnamefont {Zhu}}, \bibinfo {author} {\bibfnamefont {Y.}~\bibnamefont
  {Shi}}, \bibinfo {author} {\bibfnamefont {K.}~\bibnamefont {Jiang}}, \bibinfo
  {author} {\bibfnamefont {J.}~\bibnamefont {Hu}}, \bibinfo {author}
  {\bibfnamefont {Z.}~\bibnamefont {Xu}},\ and\ \bibinfo {author}
  {\bibfnamefont {X.~J.}\ \bibnamefont {Zhou}},\ }\bibfield  {title} {\bibinfo
  {title} {{Electronic nature of charge density wave and electron-phonon
  coupling in kagome superconductor KV$_3$Sb$_5$}},\ }\href
  {https://doi.org/10.1038/s41467-021-27946-6} {\bibfield  {journal} {\bibinfo
  {journal} {Nature Communications}\ }\textbf {\bibinfo {volume} {13}},\
  \bibinfo {pages} {273} (\bibinfo {year} {2022})}\BibitemShut {NoStop}%
\bibitem [{\citenamefont {Jiang}\ \emph
  {et~al.}(2021{\natexlab{b}})\citenamefont {Jiang}, \citenamefont {Yin},
  \citenamefont {Denner}, \citenamefont {Shumiya}, \citenamefont {Ortiz},
  \citenamefont {Xu}, \citenamefont {Guguchia}, \citenamefont {He},
  \citenamefont {Hossain}, \citenamefont {Liu}, \citenamefont {Ruff},
  \citenamefont {Kautzsch}, \citenamefont {Zhang}, \citenamefont {Chang},
  \citenamefont {Belopolski}, \citenamefont {Zhang}, \citenamefont {Cochran},
  \citenamefont {Multer}, \citenamefont {Litskevich}, \citenamefont {Cheng},
  \citenamefont {Yang}, \citenamefont {Wang}, \citenamefont {Thomale},
  \citenamefont {Neupert}, \citenamefont {Wilson},\ and\ \citenamefont
  {Hasan}}]{jiang2020discovery}%
  \BibitemOpen
  \bibfield  {author} {\bibinfo {author} {\bibfnamefont {Y.-X.}\ \bibnamefont
  {Jiang}}, \bibinfo {author} {\bibfnamefont {J.-X.}\ \bibnamefont {Yin}},
  \bibinfo {author} {\bibfnamefont {M.~M.}\ \bibnamefont {Denner}}, \bibinfo
  {author} {\bibfnamefont {N.}~\bibnamefont {Shumiya}}, \bibinfo {author}
  {\bibfnamefont {B.~R.}\ \bibnamefont {Ortiz}}, \bibinfo {author}
  {\bibfnamefont {G.}~\bibnamefont {Xu}}, \bibinfo {author} {\bibfnamefont
  {Z.}~\bibnamefont {Guguchia}}, \bibinfo {author} {\bibfnamefont
  {J.}~\bibnamefont {He}}, \bibinfo {author} {\bibfnamefont {M.~S.}\
  \bibnamefont {Hossain}}, \bibinfo {author} {\bibfnamefont {X.}~\bibnamefont
  {Liu}}, \bibinfo {author} {\bibfnamefont {J.}~\bibnamefont {Ruff}}, \bibinfo
  {author} {\bibfnamefont {L.}~\bibnamefont {Kautzsch}}, \bibinfo {author}
  {\bibfnamefont {S.~S.}\ \bibnamefont {Zhang}}, \bibinfo {author}
  {\bibfnamefont {G.}~\bibnamefont {Chang}}, \bibinfo {author} {\bibfnamefont
  {I.}~\bibnamefont {Belopolski}}, \bibinfo {author} {\bibfnamefont
  {Q.}~\bibnamefont {Zhang}}, \bibinfo {author} {\bibfnamefont {T.~A.}\
  \bibnamefont {Cochran}}, \bibinfo {author} {\bibfnamefont {D.}~\bibnamefont
  {Multer}}, \bibinfo {author} {\bibfnamefont {M.}~\bibnamefont {Litskevich}},
  \bibinfo {author} {\bibfnamefont {Z.-J.}\ \bibnamefont {Cheng}}, \bibinfo
  {author} {\bibfnamefont {X.~P.}\ \bibnamefont {Yang}}, \bibinfo {author}
  {\bibfnamefont {Z.}~\bibnamefont {Wang}}, \bibinfo {author} {\bibfnamefont
  {R.}~\bibnamefont {Thomale}}, \bibinfo {author} {\bibfnamefont
  {T.}~\bibnamefont {Neupert}}, \bibinfo {author} {\bibfnamefont {S.~D.}\
  \bibnamefont {Wilson}},\ and\ \bibinfo {author} {\bibfnamefont {M.~Z.}\
  \bibnamefont {Hasan}},\ }\bibfield  {title} {\bibinfo {title}
  {{Unconventional chiral charge order in kagome superconductor
  KV$_3$Sb$_5$}},\ }\href {https://doi.org/10.1038/s41563-021-01034-y}
  {\bibfield  {journal} {\bibinfo  {journal} {Nature Materials}\ }\textbf
  {\bibinfo {volume} {20}},\ \bibinfo {pages} {1353} (\bibinfo {year}
  {2021}{\natexlab{b}})}\BibitemShut {NoStop}%
\bibitem [{\citenamefont {Shumiya}\ \emph {et~al.}(2021)\citenamefont
  {Shumiya}, \citenamefont {Hossain}, \citenamefont {Yin}, \citenamefont
  {Jiang}, \citenamefont {Ortiz}, \citenamefont {Liu}, \citenamefont {Shi},
  \citenamefont {Yin}, \citenamefont {Lei}, \citenamefont {Zhang},
  \citenamefont {Chang}, \citenamefont {Zhang}, \citenamefont {Cochran},
  \citenamefont {Multer}, \citenamefont {Litskevich}, \citenamefont {Cheng},
  \citenamefont {Yang}, \citenamefont {Guguchia}, \citenamefont {Wilson},\ and\
  \citenamefont {Hasan}}]{Shumiya2021PRB}%
  \BibitemOpen
  \bibfield  {author} {\bibinfo {author} {\bibfnamefont {N.}~\bibnamefont
  {Shumiya}}, \bibinfo {author} {\bibfnamefont {M.~S.}\ \bibnamefont
  {Hossain}}, \bibinfo {author} {\bibfnamefont {J.-X.}\ \bibnamefont {Yin}},
  \bibinfo {author} {\bibfnamefont {Y.-X.}\ \bibnamefont {Jiang}}, \bibinfo
  {author} {\bibfnamefont {B.~R.}\ \bibnamefont {Ortiz}}, \bibinfo {author}
  {\bibfnamefont {H.}~\bibnamefont {Liu}}, \bibinfo {author} {\bibfnamefont
  {Y.}~\bibnamefont {Shi}}, \bibinfo {author} {\bibfnamefont {Q.}~\bibnamefont
  {Yin}}, \bibinfo {author} {\bibfnamefont {H.}~\bibnamefont {Lei}}, \bibinfo
  {author} {\bibfnamefont {S.~S.}\ \bibnamefont {Zhang}}, \bibinfo {author}
  {\bibfnamefont {G.}~\bibnamefont {Chang}}, \bibinfo {author} {\bibfnamefont
  {Q.}~\bibnamefont {Zhang}}, \bibinfo {author} {\bibfnamefont {T.~A.}\
  \bibnamefont {Cochran}}, \bibinfo {author} {\bibfnamefont {D.}~\bibnamefont
  {Multer}}, \bibinfo {author} {\bibfnamefont {M.}~\bibnamefont {Litskevich}},
  \bibinfo {author} {\bibfnamefont {Z.-J.}\ \bibnamefont {Cheng}}, \bibinfo
  {author} {\bibfnamefont {X.~P.}\ \bibnamefont {Yang}}, \bibinfo {author}
  {\bibfnamefont {Z.}~\bibnamefont {Guguchia}}, \bibinfo {author}
  {\bibfnamefont {S.~D.}\ \bibnamefont {Wilson}},\ and\ \bibinfo {author}
  {\bibfnamefont {M.~Z.}\ \bibnamefont {Hasan}},\ }\bibfield  {title} {\bibinfo
  {title} {{Intrinsic nature of chiral charge order in the kagome
  superconductor RbV$_{3}$Sb$_{5}$}},\ }\href
  {https://doi.org/10.1103/PhysRevB.104.035131} {\bibfield  {journal} {\bibinfo
   {journal} {Phys. Rev. B}\ }\textbf {\bibinfo {volume} {104}},\ \bibinfo
  {pages} {035131} (\bibinfo {year} {2021})}\BibitemShut {NoStop}%
\bibitem [{\citenamefont {Mielke}\ \emph {et~al.}(2022)\citenamefont {Mielke},
  \citenamefont {Das}, \citenamefont {Yin}, \citenamefont {Liu}, \citenamefont
  {Gupta}, \citenamefont {Jiang}, \citenamefont {Medarde}, \citenamefont {Wu},
  \citenamefont {Lei}, \citenamefont {Chang}, \citenamefont {Dai},
  \citenamefont {Si}, \citenamefont {Miao}, \citenamefont {Thomale},
  \citenamefont {Neupert}, \citenamefont {Shi}, \citenamefont {Khasanov},
  \citenamefont {Hasan}, \citenamefont {Luetkens},\ and\ \citenamefont
  {Guguchia}}]{mielke2021timereversal}%
  \BibitemOpen
  \bibfield  {author} {\bibinfo {author} {\bibfnamefont {C.}~\bibnamefont
  {Mielke}}, \bibinfo {author} {\bibfnamefont {D.}~\bibnamefont {Das}},
  \bibinfo {author} {\bibfnamefont {J.-X.}\ \bibnamefont {Yin}}, \bibinfo
  {author} {\bibfnamefont {H.}~\bibnamefont {Liu}}, \bibinfo {author}
  {\bibfnamefont {R.}~\bibnamefont {Gupta}}, \bibinfo {author} {\bibfnamefont
  {Y.-X.}\ \bibnamefont {Jiang}}, \bibinfo {author} {\bibfnamefont
  {M.}~\bibnamefont {Medarde}}, \bibinfo {author} {\bibfnamefont
  {X.}~\bibnamefont {Wu}}, \bibinfo {author} {\bibfnamefont {H.~C.}\
  \bibnamefont {Lei}}, \bibinfo {author} {\bibfnamefont {J.}~\bibnamefont
  {Chang}}, \bibinfo {author} {\bibfnamefont {P.}~\bibnamefont {Dai}}, \bibinfo
  {author} {\bibfnamefont {Q.}~\bibnamefont {Si}}, \bibinfo {author}
  {\bibfnamefont {H.}~\bibnamefont {Miao}}, \bibinfo {author} {\bibfnamefont
  {R.}~\bibnamefont {Thomale}}, \bibinfo {author} {\bibfnamefont
  {T.}~\bibnamefont {Neupert}}, \bibinfo {author} {\bibfnamefont
  {Y.}~\bibnamefont {Shi}}, \bibinfo {author} {\bibfnamefont {R.}~\bibnamefont
  {Khasanov}}, \bibinfo {author} {\bibfnamefont {M.~Z.}\ \bibnamefont {Hasan}},
  \bibinfo {author} {\bibfnamefont {H.}~\bibnamefont {Luetkens}},\ and\
  \bibinfo {author} {\bibfnamefont {Z.}~\bibnamefont {Guguchia}},\ }\bibfield
  {title} {\bibinfo {title} {Time-reversal symmetry-breaking charge order in a
  kagome superconductor},\ }\href {https://doi.org/10.1038/s41586-021-04327-z}
  {\bibfield  {journal} {\bibinfo  {journal} {Nature}\ }\textbf {\bibinfo
  {volume} {602}},\ \bibinfo {pages} {245} (\bibinfo {year}
  {2022})}\BibitemShut {NoStop}%
\bibitem [{\citenamefont {Yu}\ \emph {et~al.}(2021{\natexlab{a}})\citenamefont
  {Yu}, \citenamefont {Wang}, \citenamefont {Zhang}, \citenamefont {Sander},
  \citenamefont {Ni}, \citenamefont {Lu}, \citenamefont {Ma}, \citenamefont
  {Wang}, \citenamefont {Zhao}, \citenamefont {Chen}, \citenamefont {Jiang},
  \citenamefont {Zhang}, \citenamefont {Yang}, \citenamefont {Zhou},
  \citenamefont {Dong}, \citenamefont {Johnson}, \citenamefont {Graf},
  \citenamefont {Hu}, \citenamefont {Gao},\ and\ \citenamefont
  {Zhao}}]{yu2021evidence}%
  \BibitemOpen
  \bibfield  {author} {\bibinfo {author} {\bibfnamefont {L.}~\bibnamefont
  {Yu}}, \bibinfo {author} {\bibfnamefont {C.}~\bibnamefont {Wang}}, \bibinfo
  {author} {\bibfnamefont {Y.}~\bibnamefont {Zhang}}, \bibinfo {author}
  {\bibfnamefont {M.}~\bibnamefont {Sander}}, \bibinfo {author} {\bibfnamefont
  {S.}~\bibnamefont {Ni}}, \bibinfo {author} {\bibfnamefont {Z.}~\bibnamefont
  {Lu}}, \bibinfo {author} {\bibfnamefont {S.}~\bibnamefont {Ma}}, \bibinfo
  {author} {\bibfnamefont {Z.}~\bibnamefont {Wang}}, \bibinfo {author}
  {\bibfnamefont {Z.}~\bibnamefont {Zhao}}, \bibinfo {author} {\bibfnamefont
  {H.}~\bibnamefont {Chen}}, \bibinfo {author} {\bibfnamefont {K.}~\bibnamefont
  {Jiang}}, \bibinfo {author} {\bibfnamefont {Y.}~\bibnamefont {Zhang}},
  \bibinfo {author} {\bibfnamefont {H.}~\bibnamefont {Yang}}, \bibinfo {author}
  {\bibfnamefont {F.}~\bibnamefont {Zhou}}, \bibinfo {author} {\bibfnamefont
  {X.}~\bibnamefont {Dong}}, \bibinfo {author} {\bibfnamefont {S.~L.}\
  \bibnamefont {Johnson}}, \bibinfo {author} {\bibfnamefont {M.~J.}\
  \bibnamefont {Graf}}, \bibinfo {author} {\bibfnamefont {J.}~\bibnamefont
  {Hu}}, \bibinfo {author} {\bibfnamefont {H.-J.}\ \bibnamefont {Gao}},\ and\
  \bibinfo {author} {\bibfnamefont {Z.}~\bibnamefont {Zhao}},\ }\href@noop {}
  {\bibinfo {title} {{Evidence of a hidden flux phase in the topological kagome
  metal CsV$_3$Sb$_5$}}} (\bibinfo {year} {2021}{\natexlab{a}}),\ \Eprint
  {https://arxiv.org/abs/2107.10714} {arXiv:2107.10714 [cond-mat.supr-con]}
  \BibitemShut {NoStop}%
\bibitem [{\citenamefont {Yang}\ \emph {et~al.}(2020)\citenamefont {Yang},
  \citenamefont {Wang}, \citenamefont {Ortiz}, \citenamefont {Liu},
  \citenamefont {Gayles}, \citenamefont {Derunova}, \citenamefont
  {Gonzalez-Hernandez}, \citenamefont {Šmejkal}, \citenamefont {Chen},
  \citenamefont {Parkin}, \citenamefont {Wilson}, \citenamefont {Toberer},
  \citenamefont {McQueen},\ and\ \citenamefont {Ali}}]{Yang2020Hall}%
  \BibitemOpen
  \bibfield  {author} {\bibinfo {author} {\bibfnamefont {S.-Y.}\ \bibnamefont
  {Yang}}, \bibinfo {author} {\bibfnamefont {Y.}~\bibnamefont {Wang}}, \bibinfo
  {author} {\bibfnamefont {B.~R.}\ \bibnamefont {Ortiz}}, \bibinfo {author}
  {\bibfnamefont {D.}~\bibnamefont {Liu}}, \bibinfo {author} {\bibfnamefont
  {J.}~\bibnamefont {Gayles}}, \bibinfo {author} {\bibfnamefont
  {E.}~\bibnamefont {Derunova}}, \bibinfo {author} {\bibfnamefont
  {R.}~\bibnamefont {Gonzalez-Hernandez}}, \bibinfo {author} {\bibfnamefont
  {L.}~\bibnamefont {Šmejkal}}, \bibinfo {author} {\bibfnamefont
  {Y.}~\bibnamefont {Chen}}, \bibinfo {author} {\bibfnamefont {S.~S.~P.}\
  \bibnamefont {Parkin}}, \bibinfo {author} {\bibfnamefont {S.~D.}\
  \bibnamefont {Wilson}}, \bibinfo {author} {\bibfnamefont {E.~S.}\
  \bibnamefont {Toberer}}, \bibinfo {author} {\bibfnamefont {T.}~\bibnamefont
  {McQueen}},\ and\ \bibinfo {author} {\bibfnamefont {M.~N.}\ \bibnamefont
  {Ali}},\ }\bibfield  {title} {\bibinfo {title} {{Giant, unconventional
  anomalous Hall effect in the metallic frustrated magnet candidate,
  KV$_{3}$Sb$_{5}$}},\ }\href {https://doi.org/10.1126/sciadv.abb6003}
  {\bibfield  {journal} {\bibinfo  {journal} {Science Advances}\ }\textbf
  {\bibinfo {volume} {6}},\ \bibinfo {pages} {eabb6003} (\bibinfo {year}
  {2020})}\BibitemShut {NoStop}%
\bibitem [{\citenamefont {Yu}\ \emph {et~al.}(2021{\natexlab{b}})\citenamefont
  {Yu}, \citenamefont {Wu}, \citenamefont {Wang}, \citenamefont {Lei},
  \citenamefont {Zhuo}, \citenamefont {Ying},\ and\ \citenamefont
  {Chen}}]{yu2021concurrence}%
  \BibitemOpen
  \bibfield  {author} {\bibinfo {author} {\bibfnamefont {F.~H.}\ \bibnamefont
  {Yu}}, \bibinfo {author} {\bibfnamefont {T.}~\bibnamefont {Wu}}, \bibinfo
  {author} {\bibfnamefont {Z.~Y.}\ \bibnamefont {Wang}}, \bibinfo {author}
  {\bibfnamefont {B.}~\bibnamefont {Lei}}, \bibinfo {author} {\bibfnamefont
  {W.~Z.}\ \bibnamefont {Zhuo}}, \bibinfo {author} {\bibfnamefont {J.~J.}\
  \bibnamefont {Ying}},\ and\ \bibinfo {author} {\bibfnamefont {X.~H.}\
  \bibnamefont {Chen}},\ }\bibfield  {title} {\bibinfo {title} {{Concurrence of
  anomalous Hall effect and charge density wave in a superconducting
  topological kagome metal}},\ }\href
  {https://doi.org/10.1103/PhysRevB.104.L041103} {\bibfield  {journal}
  {\bibinfo  {journal} {Phys. Rev. B}\ }\textbf {\bibinfo {volume} {104}},\
  \bibinfo {pages} {L041103} (\bibinfo {year}
  {2021}{\natexlab{b}})}\BibitemShut {NoStop}%
\bibitem [{\citenamefont {Nagaosa}\ \emph {et~al.}(2010)\citenamefont
  {Nagaosa}, \citenamefont {Sinova}, \citenamefont {Onoda}, \citenamefont
  {MacDonald},\ and\ \citenamefont {Ong}}]{Nagaosa2010REVIEWSOFMODERNPHYSICS}%
  \BibitemOpen
  \bibfield  {author} {\bibinfo {author} {\bibfnamefont {N.}~\bibnamefont
  {Nagaosa}}, \bibinfo {author} {\bibfnamefont {J.}~\bibnamefont {Sinova}},
  \bibinfo {author} {\bibfnamefont {S.}~\bibnamefont {Onoda}}, \bibinfo
  {author} {\bibfnamefont {A.~H.}\ \bibnamefont {MacDonald}},\ and\ \bibinfo
  {author} {\bibfnamefont {N.~P.}\ \bibnamefont {Ong}},\ }\bibfield  {title}
  {\bibinfo {title} {{Anomalous Hall effect}},\ }\href
  {https://doi.org/10.1103/RevModPhys.82.1539} {\bibfield  {journal} {\bibinfo
  {journal} {Rev. Mod. Phys.}\ }\textbf {\bibinfo {volume} {82}},\ \bibinfo
  {pages} {1539} (\bibinfo {year} {2010})}\BibitemShut {NoStop}%
\bibitem [{\citenamefont {Xiao}\ \emph {et~al.}(2010)\citenamefont {Xiao},
  \citenamefont {Chang},\ and\ \citenamefont
  {Niu}}]{XiaoDi2010REVIEWSOFMODERNPHYSICS}%
  \BibitemOpen
  \bibfield  {author} {\bibinfo {author} {\bibfnamefont {D.}~\bibnamefont
  {Xiao}}, \bibinfo {author} {\bibfnamefont {M.-C.}\ \bibnamefont {Chang}},\
  and\ \bibinfo {author} {\bibfnamefont {Q.}~\bibnamefont {Niu}},\ }\bibfield
  {title} {\bibinfo {title} {{Berry phase effects on electronic properties}},\
  }\href {https://doi.org/10.1103/RevModPhys.82.1959} {\bibfield  {journal}
  {\bibinfo  {journal} {Rev. Mod. Phys.}\ }\textbf {\bibinfo {volume} {82}},\
  \bibinfo {pages} {1959} (\bibinfo {year} {2010})}\BibitemShut {NoStop}%
\bibitem [{\citenamefont {Li}\ and\ \citenamefont {Chen}(2020)}]{Li2020MRS}%
  \BibitemOpen
  \bibfield  {author} {\bibinfo {author} {\bibfnamefont {M.}~\bibnamefont
  {Li}}\ and\ \bibinfo {author} {\bibfnamefont {G.}~\bibnamefont {Chen}},\
  }\bibfield  {title} {\bibinfo {title} {Thermal transport for probing quantum
  materials},\ }\href {https://doi.org/10.1557/mrs.2020.124} {\bibfield
  {journal} {\bibinfo  {journal} {MRS Bulletin}\ }\textbf {\bibinfo {volume}
  {45}},\ \bibinfo {pages} {348} (\bibinfo {year} {2020})}\BibitemShut
  {NoStop}%
\bibitem [{\citenamefont {Xiao}\ \emph {et~al.}(2006)\citenamefont {Xiao},
  \citenamefont {Yao}, \citenamefont {Fang},\ and\ \citenamefont
  {Niu}}]{XiaoDi2006PHYSICALREVIEWLETTERS}%
  \BibitemOpen
  \bibfield  {author} {\bibinfo {author} {\bibfnamefont {D.}~\bibnamefont
  {Xiao}}, \bibinfo {author} {\bibfnamefont {Y.}~\bibnamefont {Yao}}, \bibinfo
  {author} {\bibfnamefont {Z.}~\bibnamefont {Fang}},\ and\ \bibinfo {author}
  {\bibfnamefont {Q.}~\bibnamefont {Niu}},\ }\bibfield  {title} {\bibinfo
  {title} {Berry-phase effect in anomalous thermoelectric transport},\ }\href
  {https://doi.org/10.1103/PhysRevLett.97.026603} {\bibfield  {journal}
  {\bibinfo  {journal} {Phys. Rev. Lett.}\ }\textbf {\bibinfo {volume} {97}},\
  \bibinfo {pages} {026603} (\bibinfo {year} {2006})}\BibitemShut {NoStop}%
\bibitem [{\citenamefont {Onoda}\ \emph {et~al.}(2008)\citenamefont {Onoda},
  \citenamefont {Sugimoto},\ and\ \citenamefont
  {Nagaosa}}]{OnodaS2008PhysicalReviewB}%
  \BibitemOpen
  \bibfield  {author} {\bibinfo {author} {\bibfnamefont {S.}~\bibnamefont
  {Onoda}}, \bibinfo {author} {\bibfnamefont {N.}~\bibnamefont {Sugimoto}},\
  and\ \bibinfo {author} {\bibfnamefont {N.}~\bibnamefont {Nagaosa}},\
  }\bibfield  {title} {\bibinfo {title} {{Quantum transport theory of anomalous
  electric, thermoelectric, and thermal Hall effects in ferromagnets}},\ }\href
  {https://doi.org/10.1103/PhysRevB.77.165103} {\bibfield  {journal} {\bibinfo
  {journal} {Phys. Rev. B}\ }\textbf {\bibinfo {volume} {77}},\ \bibinfo
  {pages} {165103} (\bibinfo {year} {2008})}\BibitemShut {NoStop}%
\bibitem [{\citenamefont {Shiomi}\ \emph {et~al.}(2010)\citenamefont {Shiomi},
  \citenamefont {Onose},\ and\ \citenamefont
  {Tokura}}]{ShiomiY2010PhysicalReviewB}%
  \BibitemOpen
  \bibfield  {author} {\bibinfo {author} {\bibfnamefont {Y.}~\bibnamefont
  {Shiomi}}, \bibinfo {author} {\bibfnamefont {Y.}~\bibnamefont {Onose}},\ and\
  \bibinfo {author} {\bibfnamefont {Y.}~\bibnamefont {Tokura}},\ }\bibfield
  {title} {\bibinfo {title} {{Effect of scattering on intrinsic anomalous Hall
  effect investigated by Lorenz ratio}},\ }\href
  {https://doi.org/10.1103/PhysRevB.81.054414} {\bibfield  {journal} {\bibinfo
  {journal} {Phys. Rev. B}\ }\textbf {\bibinfo {volume} {81}},\ \bibinfo
  {pages} {054414} (\bibinfo {year} {2010})}\BibitemShut {NoStop}%
\bibitem [{\citenamefont {Li}\ \emph {et~al.}(2017)\citenamefont {Li},
  \citenamefont {Xu}, \citenamefont {Ding}, \citenamefont {Wang}, \citenamefont
  {Shen}, \citenamefont {Lu}, \citenamefont {Zhu},\ and\ \citenamefont
  {Behnia}}]{LiXiaokang2017PhysRevLett}%
  \BibitemOpen
  \bibfield  {author} {\bibinfo {author} {\bibfnamefont {X.}~\bibnamefont
  {Li}}, \bibinfo {author} {\bibfnamefont {L.}~\bibnamefont {Xu}}, \bibinfo
  {author} {\bibfnamefont {L.}~\bibnamefont {Ding}}, \bibinfo {author}
  {\bibfnamefont {J.}~\bibnamefont {Wang}}, \bibinfo {author} {\bibfnamefont
  {M.}~\bibnamefont {Shen}}, \bibinfo {author} {\bibfnamefont {X.}~\bibnamefont
  {Lu}}, \bibinfo {author} {\bibfnamefont {Z.}~\bibnamefont {Zhu}},\ and\
  \bibinfo {author} {\bibfnamefont {K.}~\bibnamefont {Behnia}},\ }\bibfield
  {title} {\bibinfo {title} {{Anomalous Nernst and Righi-Leduc Effects in
  ${\mathrm{Mn}}_{3}\mathrm{Sn}$: Berry Curvature and Entropy Flow}},\ }\href
  {https://doi.org/10.1103/PhysRevLett.119.056601} {\bibfield  {journal}
  {\bibinfo  {journal} {Phys. Rev. Lett.}\ }\textbf {\bibinfo {volume} {119}},\
  \bibinfo {pages} {056601} (\bibinfo {year} {2017})}\BibitemShut {NoStop}%
\bibitem [{\citenamefont {Xu}\ \emph {et~al.}(2020{\natexlab{a}})\citenamefont
  {Xu}, \citenamefont {Li}, \citenamefont {Lu}, \citenamefont {Collignon},
  \citenamefont {Fu}, \citenamefont {Koo}, \citenamefont {Fauqu{\'e}},
  \citenamefont {Yan}, \citenamefont {Zhu},\ and\ \citenamefont
  {Behnia}}]{ZhuZengwei2020}%
  \BibitemOpen
  \bibfield  {author} {\bibinfo {author} {\bibfnamefont {L.}~\bibnamefont
  {Xu}}, \bibinfo {author} {\bibfnamefont {X.}~\bibnamefont {Li}}, \bibinfo
  {author} {\bibfnamefont {X.}~\bibnamefont {Lu}}, \bibinfo {author}
  {\bibfnamefont {C.}~\bibnamefont {Collignon}}, \bibinfo {author}
  {\bibfnamefont {H.}~\bibnamefont {Fu}}, \bibinfo {author} {\bibfnamefont
  {J.}~\bibnamefont {Koo}}, \bibinfo {author} {\bibfnamefont {B.}~\bibnamefont
  {Fauqu{\'e}}}, \bibinfo {author} {\bibfnamefont {B.}~\bibnamefont {Yan}},
  \bibinfo {author} {\bibfnamefont {Z.}~\bibnamefont {Zhu}},\ and\ \bibinfo
  {author} {\bibfnamefont {K.}~\bibnamefont {Behnia}},\ }\bibfield  {title}
  {\bibinfo {title} {{Finite-temperature violation of the anomalous transverse
  Wiedemann-Franz law}},\ }\href {https://doi.org/10.1126/sciadv.aaz3522}
  {\bibfield  {journal} {\bibinfo  {journal} {Science Advances}\ }\textbf
  {\bibinfo {volume} {6}},\ \bibinfo {pages} {eaaz3522} (\bibinfo {year}
  {2020}{\natexlab{a}})}\BibitemShut {NoStop}%
\bibitem [{\citenamefont {Grissonnanche}\ \emph {et~al.}(2019)\citenamefont
  {Grissonnanche}, \citenamefont {Legros}, \citenamefont {Badoux},
  \citenamefont {Lefran{\c{c}}ois}, \citenamefont {Zatko}, \citenamefont
  {Lizaire}, \citenamefont {Lalibert{\'e}}, \citenamefont {Gourgout},
  \citenamefont {Zhou}, \citenamefont {Pyon}, \citenamefont {Takayama},
  \citenamefont {Takagi}, \citenamefont {Ono}, \citenamefont {Doiron-Leyraud},\
  and\ \citenamefont {Taillefer}}]{GrissonnancheG2019NATURE}%
  \BibitemOpen
  \bibfield  {author} {\bibinfo {author} {\bibfnamefont {G.}~\bibnamefont
  {Grissonnanche}}, \bibinfo {author} {\bibfnamefont {A.}~\bibnamefont
  {Legros}}, \bibinfo {author} {\bibfnamefont {S.}~\bibnamefont {Badoux}},
  \bibinfo {author} {\bibfnamefont {E.}~\bibnamefont {Lefran{\c{c}}ois}},
  \bibinfo {author} {\bibfnamefont {V.}~\bibnamefont {Zatko}}, \bibinfo
  {author} {\bibfnamefont {M.}~\bibnamefont {Lizaire}}, \bibinfo {author}
  {\bibfnamefont {F.}~\bibnamefont {Lalibert{\'e}}}, \bibinfo {author}
  {\bibfnamefont {A.}~\bibnamefont {Gourgout}}, \bibinfo {author}
  {\bibfnamefont {J.-S.}\ \bibnamefont {Zhou}}, \bibinfo {author}
  {\bibfnamefont {S.}~\bibnamefont {Pyon}}, \bibinfo {author} {\bibfnamefont
  {T.}~\bibnamefont {Takayama}}, \bibinfo {author} {\bibfnamefont
  {H.}~\bibnamefont {Takagi}}, \bibinfo {author} {\bibfnamefont
  {S.}~\bibnamefont {Ono}}, \bibinfo {author} {\bibfnamefont {N.}~\bibnamefont
  {Doiron-Leyraud}},\ and\ \bibinfo {author} {\bibfnamefont {L.}~\bibnamefont
  {Taillefer}},\ }\bibfield  {title} {\bibinfo {title} {Giant thermal hall
  conductivity in the pseudogap phase of cuprate superconductors},\ }\href
  {https://doi.org/10.1038/s41586-019-1375-0} {\bibfield  {journal} {\bibinfo
  {journal} {Nature}\ }\textbf {\bibinfo {volume} {571}},\ \bibinfo {pages}
  {376} (\bibinfo {year} {2019})}\BibitemShut {NoStop}%
\bibitem [{\citenamefont {Hirschberger}\ \emph {et~al.}(2015)\citenamefont
  {Hirschberger}, \citenamefont {Krizan}, \citenamefont {Cava},\ and\
  \citenamefont {Ong}}]{Hirschberger2015SCIENCE}%
  \BibitemOpen
  \bibfield  {author} {\bibinfo {author} {\bibfnamefont {M.}~\bibnamefont
  {Hirschberger}}, \bibinfo {author} {\bibfnamefont {J.~W.}\ \bibnamefont
  {Krizan}}, \bibinfo {author} {\bibfnamefont {R.~J.}\ \bibnamefont {Cava}},\
  and\ \bibinfo {author} {\bibfnamefont {N.~P.}\ \bibnamefont {Ong}},\
  }\bibfield  {title} {\bibinfo {title} {{Large thermal Hall conductivity of
  neutral spin excitations in a frustrated quantum magnet}},\ }\href
  {https://doi.org/10.1126/science.1257340} {\bibfield  {journal} {\bibinfo
  {journal} {Science}\ }\textbf {\bibinfo {volume} {348}},\ \bibinfo {pages}
  {106} (\bibinfo {year} {2015})}\BibitemShut {NoStop}%
\bibitem [{\citenamefont {Onose}\ \emph {et~al.}(2010)\citenamefont {Onose},
  \citenamefont {Ideue}, \citenamefont {Katsura}, \citenamefont {Shiomi},
  \citenamefont {Nagaosa},\ and\ \citenamefont {Tokura}}]{OnoseY2010SCIENCE}%
  \BibitemOpen
  \bibfield  {author} {\bibinfo {author} {\bibfnamefont {Y.}~\bibnamefont
  {Onose}}, \bibinfo {author} {\bibfnamefont {T.}~\bibnamefont {Ideue}},
  \bibinfo {author} {\bibfnamefont {H.}~\bibnamefont {Katsura}}, \bibinfo
  {author} {\bibfnamefont {Y.}~\bibnamefont {Shiomi}}, \bibinfo {author}
  {\bibfnamefont {N.}~\bibnamefont {Nagaosa}},\ and\ \bibinfo {author}
  {\bibfnamefont {Y.}~\bibnamefont {Tokura}},\ }\bibfield  {title} {\bibinfo
  {title} {{Observation of the Magnon Hall Effect}},\ }\href
  {https://doi.org/10.1126/science.1188260} {\bibfield  {journal} {\bibinfo
  {journal} {Science}\ }\textbf {\bibinfo {volume} {329}},\ \bibinfo {pages}
  {297} (\bibinfo {year} {2010})}\BibitemShut {NoStop}%
\bibitem [{\citenamefont {Strohm}\ \emph {et~al.}(2005)\citenamefont {Strohm},
  \citenamefont {Rikken},\ and\ \citenamefont {Wyder}}]{Strohm2005prl}%
  \BibitemOpen
  \bibfield  {author} {\bibinfo {author} {\bibfnamefont {C.}~\bibnamefont
  {Strohm}}, \bibinfo {author} {\bibfnamefont {G.~L. J.~A.}\ \bibnamefont
  {Rikken}},\ and\ \bibinfo {author} {\bibfnamefont {P.}~\bibnamefont
  {Wyder}},\ }\bibfield  {title} {\bibinfo {title} {Phenomenological evidence
  for the phonon hall effect},\ }\href
  {https://doi.org/10.1103/PhysRevLett.95.155901} {\bibfield  {journal}
  {\bibinfo  {journal} {Phys. Rev. Lett.}\ }\textbf {\bibinfo {volume} {95}},\
  \bibinfo {pages} {155901} (\bibinfo {year} {2005})}\BibitemShut {NoStop}%
\bibitem [{\citenamefont {Sheng}\ \emph {et~al.}(2006)\citenamefont {Sheng},
  \citenamefont {Sheng},\ and\ \citenamefont {Ting}}]{Sheng2006prl}%
  \BibitemOpen
  \bibfield  {author} {\bibinfo {author} {\bibfnamefont {L.}~\bibnamefont
  {Sheng}}, \bibinfo {author} {\bibfnamefont {D.~N.}\ \bibnamefont {Sheng}},\
  and\ \bibinfo {author} {\bibfnamefont {C.~S.}\ \bibnamefont {Ting}},\
  }\bibfield  {title} {\bibinfo {title} {Theory of the phonon hall effect in
  paramagnetic dielectrics},\ }\href
  {https://doi.org/10.1103/PhysRevLett.96.155901} {\bibfield  {journal}
  {\bibinfo  {journal} {Phys. Rev. Lett.}\ }\textbf {\bibinfo {volume} {96}},\
  \bibinfo {pages} {155901} (\bibinfo {year} {2006})}\BibitemShut {NoStop}%
\bibitem [{\citenamefont {Kagan}\ and\ \citenamefont
  {Maksimov}(2008)}]{Kagan2008prl}%
  \BibitemOpen
  \bibfield  {author} {\bibinfo {author} {\bibfnamefont {Y.}~\bibnamefont
  {Kagan}}\ and\ \bibinfo {author} {\bibfnamefont {L.~A.}\ \bibnamefont
  {Maksimov}},\ }\bibfield  {title} {\bibinfo {title} {Anomalous hall effect
  for the phonon heat conductivity in paramagnetic dielectrics},\ }\href
  {https://doi.org/10.1103/PhysRevLett.100.145902} {\bibfield  {journal}
  {\bibinfo  {journal} {Phys. Rev. Lett.}\ }\textbf {\bibinfo {volume} {100}},\
  \bibinfo {pages} {145902} (\bibinfo {year} {2008})}\BibitemShut {NoStop}%
\bibitem [{\citenamefont {Mori}\ \emph {et~al.}(2014)\citenamefont {Mori},
  \citenamefont {Spencer-Smith}, \citenamefont {Sushkov},\ and\ \citenamefont
  {Maekawa}}]{Mori2014prl}%
  \BibitemOpen
  \bibfield  {author} {\bibinfo {author} {\bibfnamefont {M.}~\bibnamefont
  {Mori}}, \bibinfo {author} {\bibfnamefont {A.}~\bibnamefont {Spencer-Smith}},
  \bibinfo {author} {\bibfnamefont {O.~P.}\ \bibnamefont {Sushkov}},\ and\
  \bibinfo {author} {\bibfnamefont {S.}~\bibnamefont {Maekawa}},\ }\bibfield
  {title} {\bibinfo {title} {Origin of the phonon hall effect in rare-earth
  garnets},\ }\href {https://doi.org/10.1103/PhysRevLett.113.265901} {\bibfield
   {journal} {\bibinfo  {journal} {Phys. Rev. Lett.}\ }\textbf {\bibinfo
  {volume} {113}},\ \bibinfo {pages} {265901} (\bibinfo {year}
  {2014})}\BibitemShut {NoStop}%
\bibitem [{\citenamefont {Zheng}\ \emph {et~al.}(2021)\citenamefont {Zheng},
  \citenamefont {Chen}, \citenamefont {Tan}, \citenamefont {Wang},
  \citenamefont {Zhu}, \citenamefont {Albarakati}, \citenamefont {Algarni},
  \citenamefont {Partridge}, \citenamefont {Farrar}, \citenamefont {Zhou},
  \citenamefont {Ning}, \citenamefont {Tian}, \citenamefont {Fuhrer},\ and\
  \citenamefont {Wang}}]{Guolin2021arxiv}%
  \BibitemOpen
  \bibfield  {author} {\bibinfo {author} {\bibfnamefont {G.}~\bibnamefont
  {Zheng}}, \bibinfo {author} {\bibfnamefont {Z.}~\bibnamefont {Chen}},
  \bibinfo {author} {\bibfnamefont {C.}~\bibnamefont {Tan}}, \bibinfo {author}
  {\bibfnamefont {M.}~\bibnamefont {Wang}}, \bibinfo {author} {\bibfnamefont
  {X.}~\bibnamefont {Zhu}}, \bibinfo {author} {\bibfnamefont {S.}~\bibnamefont
  {Albarakati}}, \bibinfo {author} {\bibfnamefont {M.}~\bibnamefont {Algarni}},
  \bibinfo {author} {\bibfnamefont {J.}~\bibnamefont {Partridge}}, \bibinfo
  {author} {\bibfnamefont {L.}~\bibnamefont {Farrar}}, \bibinfo {author}
  {\bibfnamefont {J.}~\bibnamefont {Zhou}}, \bibinfo {author} {\bibfnamefont
  {W.}~\bibnamefont {Ning}}, \bibinfo {author} {\bibfnamefont {M.}~\bibnamefont
  {Tian}}, \bibinfo {author} {\bibfnamefont {M.~S.}\ \bibnamefont {Fuhrer}},\
  and\ \bibinfo {author} {\bibfnamefont {L.}~\bibnamefont {Wang}},\ }\href@noop
  {} {\bibinfo {title} {{Gate-controllable giant anomalous Hall effect from
  flat bands in kagome metal CsV$_3$Sb$_5$ nanoflakes}}} (\bibinfo {year}
  {2021}),\ \Eprint {https://arxiv.org/abs/2109.12588} {arXiv:2109.12588
  [cond-mat.mtrl-sci]} \BibitemShut {NoStop}%
\bibitem [{\citenamefont {Behnia}(2009)}]{Behnia2009}%
  \BibitemOpen
  \bibfield  {author} {\bibinfo {author} {\bibfnamefont {K.}~\bibnamefont
  {Behnia}},\ }\bibfield  {title} {\bibinfo {title} {The nernst effect and the
  boundaries of the fermi liquid picture},\ }\href
  {https://doi.org/10.1088/0953-8984/21/11/113101} {\bibfield  {journal}
  {\bibinfo  {journal} {Journal of Physics: Condensed Matter}\ }\textbf
  {\bibinfo {volume} {21}},\ \bibinfo {pages} {113101} (\bibinfo {year}
  {2009})}\BibitemShut {NoStop}%
\bibitem [{\citenamefont {Kim}\ \emph {et~al.}(2012)\citenamefont {Kim},
  \citenamefont {Rhyee},\ and\ \citenamefont {Kwon}}]{Kim2012PhysRevB}%
  \BibitemOpen
  \bibfield  {author} {\bibinfo {author} {\bibfnamefont {J.~H.}\ \bibnamefont
  {Kim}}, \bibinfo {author} {\bibfnamefont {J.-S.}\ \bibnamefont {Rhyee}},\
  and\ \bibinfo {author} {\bibfnamefont {Y.~S.}\ \bibnamefont {Kwon}},\
  }\bibfield  {title} {\bibinfo {title} {{Magnon gap formation and charge
  density wave effect on thermoelectric properties in the SmNiC${}_{2}$
  compound}},\ }\href {https://doi.org/10.1103/PhysRevB.86.235101} {\bibfield
  {journal} {\bibinfo  {journal} {Phys. Rev. B}\ }\textbf {\bibinfo {volume}
  {86}},\ \bibinfo {pages} {235101} (\bibinfo {year} {2012})}\BibitemShut
  {NoStop}%
\bibitem [{\citenamefont {Kuo}\ \emph {et~al.}(2006)\citenamefont {Kuo},
  \citenamefont {Sivakumar}, \citenamefont {Su},\ and\ \citenamefont
  {Lue}}]{Kuo2006PhysRevB}%
  \BibitemOpen
  \bibfield  {author} {\bibinfo {author} {\bibfnamefont {Y.~K.}\ \bibnamefont
  {Kuo}}, \bibinfo {author} {\bibfnamefont {K.~M.}\ \bibnamefont {Sivakumar}},
  \bibinfo {author} {\bibfnamefont {T.~H.}\ \bibnamefont {Su}},\ and\ \bibinfo
  {author} {\bibfnamefont {C.~S.}\ \bibnamefont {Lue}},\ }\bibfield  {title}
  {\bibinfo {title} {{Phase transitions in
  ${\mathrm{Lu}}_{2}{\mathrm{Ir}}_{3}{\mathrm{Si}}_{5}$: An experimental
  investigation by transport measurements}},\ }\href
  {https://doi.org/10.1103/PhysRevB.74.045115} {\bibfield  {journal} {\bibinfo
  {journal} {Phys. Rev. B}\ }\textbf {\bibinfo {volume} {74}},\ \bibinfo
  {pages} {045115} (\bibinfo {year} {2006})}\BibitemShut {NoStop}%
\bibitem [{\citenamefont {Gumeniuk}\ \emph {et~al.}(2015)\citenamefont
  {Gumeniuk}, \citenamefont {Kvashnina}, \citenamefont {Schnelle},
  \citenamefont {Leithe-Jasper},\ and\ \citenamefont
  {Grin}}]{Gumeniuk2015PhysRevB}%
  \BibitemOpen
  \bibfield  {author} {\bibinfo {author} {\bibfnamefont {R.}~\bibnamefont
  {Gumeniuk}}, \bibinfo {author} {\bibfnamefont {K.~O.}\ \bibnamefont
  {Kvashnina}}, \bibinfo {author} {\bibfnamefont {W.}~\bibnamefont {Schnelle}},
  \bibinfo {author} {\bibfnamefont {A.}~\bibnamefont {Leithe-Jasper}},\ and\
  \bibinfo {author} {\bibfnamefont {Y.}~\bibnamefont {Grin}},\ }\bibfield
  {title} {\bibinfo {title} {{Magnetic and transport properties of structural
  variants of Remeika phases:
  ${\mathrm{Th}}_{3}{\mathrm{Ir}}_{4}{\mathrm{Ge}}_{13}$ and
  ${\mathrm{U}}_{3}{\mathrm{Ir}}_{4}{\mathrm{Ge}}_{13}$}},\ }\href
  {https://doi.org/10.1103/PhysRevB.91.094110} {\bibfield  {journal} {\bibinfo
  {journal} {Phys. Rev. B}\ }\textbf {\bibinfo {volume} {91}},\ \bibinfo
  {pages} {094110} (\bibinfo {year} {2015})}\BibitemShut {NoStop}%
\bibitem [{\citenamefont {Kuo}\ \emph {et~al.}(2020)\citenamefont {Kuo},
  \citenamefont {Huang}, \citenamefont {Kuo},\ and\ \citenamefont
  {Lue}}]{Kuo2020PhysRevB}%
  \BibitemOpen
  \bibfield  {author} {\bibinfo {author} {\bibfnamefont {C.~N.}\ \bibnamefont
  {Kuo}}, \bibinfo {author} {\bibfnamefont {R.~Y.}\ \bibnamefont {Huang}},
  \bibinfo {author} {\bibfnamefont {Y.~K.}\ \bibnamefont {Kuo}},\ and\ \bibinfo
  {author} {\bibfnamefont {C.~S.}\ \bibnamefont {Lue}},\ }\bibfield  {title}
  {\bibinfo {title} {{Transport and thermal behavior of the charge density wave
  phase transition in CuTe}},\ }\href
  {https://doi.org/10.1103/PhysRevB.102.155137} {\bibfield  {journal} {\bibinfo
   {journal} {Phys. Rev. B}\ }\textbf {\bibinfo {volume} {102}},\ \bibinfo
  {pages} {155137} (\bibinfo {year} {2020})}\BibitemShut {NoStop}%
\bibitem [{\citenamefont {Kountz}\ \emph {et~al.}(2021)\citenamefont {Kountz},
  \citenamefont {Zhang}, \citenamefont {Straquadine}, \citenamefont {Singh},
  \citenamefont {Bachmann}, \citenamefont {Fisher}, \citenamefont {Kivelson},\
  and\ \citenamefont {Kapitulnik}}]{Kountz2021PhysRevB}%
  \BibitemOpen
  \bibfield  {author} {\bibinfo {author} {\bibfnamefont {E.~D.}\ \bibnamefont
  {Kountz}}, \bibinfo {author} {\bibfnamefont {J.}~\bibnamefont {Zhang}},
  \bibinfo {author} {\bibfnamefont {J.~A.~W.}\ \bibnamefont {Straquadine}},
  \bibinfo {author} {\bibfnamefont {A.~G.}\ \bibnamefont {Singh}}, \bibinfo
  {author} {\bibfnamefont {M.~D.}\ \bibnamefont {Bachmann}}, \bibinfo {author}
  {\bibfnamefont {I.~R.}\ \bibnamefont {Fisher}}, \bibinfo {author}
  {\bibfnamefont {S.~A.}\ \bibnamefont {Kivelson}},\ and\ \bibinfo {author}
  {\bibfnamefont {A.}~\bibnamefont {Kapitulnik}},\ }\bibfield  {title}
  {\bibinfo {title} {{Anomalous thermal transport and strong violation of
  Wiedemann-Franz law in the critical regime of a charge density wave
  transition}},\ }\href {https://doi.org/10.1103/PhysRevB.104.L241109}
  {\bibfield  {journal} {\bibinfo  {journal} {Phys. Rev. B}\ }\textbf {\bibinfo
  {volume} {104}},\ \bibinfo {pages} {L241109} (\bibinfo {year}
  {2021})}\BibitemShut {NoStop}%
\bibitem [{\citenamefont {Li}\ \emph {et~al.}(2021{\natexlab{a}})\citenamefont
  {Li}, \citenamefont {Zhang}, \citenamefont {Yilmaz}, \citenamefont {Pai},
  \citenamefont {Marvinney}, \citenamefont {Said}, \citenamefont {Yin},
  \citenamefont {Gong}, \citenamefont {Tu}, \citenamefont {Vescovo},
  \citenamefont {Nelson}, \citenamefont {Moore}, \citenamefont {Murakami},
  \citenamefont {Lei}, \citenamefont {Lee}, \citenamefont {Lawrie},\ and\
  \citenamefont {Miao}}]{LiHaoxiang2021prx}%
  \BibitemOpen
  \bibfield  {author} {\bibinfo {author} {\bibfnamefont {H.}~\bibnamefont
  {Li}}, \bibinfo {author} {\bibfnamefont {T.~T.}\ \bibnamefont {Zhang}},
  \bibinfo {author} {\bibfnamefont {T.}~\bibnamefont {Yilmaz}}, \bibinfo
  {author} {\bibfnamefont {Y.~Y.}\ \bibnamefont {Pai}}, \bibinfo {author}
  {\bibfnamefont {C.~E.}\ \bibnamefont {Marvinney}}, \bibinfo {author}
  {\bibfnamefont {A.}~\bibnamefont {Said}}, \bibinfo {author} {\bibfnamefont
  {Q.~W.}\ \bibnamefont {Yin}}, \bibinfo {author} {\bibfnamefont {C.~S.}\
  \bibnamefont {Gong}}, \bibinfo {author} {\bibfnamefont {Z.~J.}\ \bibnamefont
  {Tu}}, \bibinfo {author} {\bibfnamefont {E.}~\bibnamefont {Vescovo}},
  \bibinfo {author} {\bibfnamefont {C.~S.}\ \bibnamefont {Nelson}}, \bibinfo
  {author} {\bibfnamefont {R.~G.}\ \bibnamefont {Moore}}, \bibinfo {author}
  {\bibfnamefont {S.}~\bibnamefont {Murakami}}, \bibinfo {author}
  {\bibfnamefont {H.~C.}\ \bibnamefont {Lei}}, \bibinfo {author} {\bibfnamefont
  {H.~N.}\ \bibnamefont {Lee}}, \bibinfo {author} {\bibfnamefont {B.~J.}\
  \bibnamefont {Lawrie}},\ and\ \bibinfo {author} {\bibfnamefont
  {H.}~\bibnamefont {Miao}},\ }\bibfield  {title} {\bibinfo {title}
  {{Observation of Unconventional Charge Density Wave without Acoustic Phonon
  Anomaly in Kagome Superconductors ${A\mathrm{V}}_{3}{\mathrm{Sb}}_{5}$
  ($A=\mathrm{Rb}, \mathrm{Cs}$)}},\ }\href
  {https://doi.org/10.1103/PhysRevX.11.031050} {\bibfield  {journal} {\bibinfo
  {journal} {Phys. Rev. X}\ }\textbf {\bibinfo {volume} {11}},\ \bibinfo
  {pages} {031050} (\bibinfo {year} {2021}{\natexlab{a}})}\BibitemShut
  {NoStop}%
\bibitem [{\citenamefont {Wang}\ \emph {et~al.}(2021)\citenamefont {Wang},
  \citenamefont {Wu}, \citenamefont {Yin}, \citenamefont {Gong}, \citenamefont
  {Tu}, \citenamefont {Lin}, \citenamefont {Liu}, \citenamefont {Shi},
  \citenamefont {Zhang}, \citenamefont {Wu}, \citenamefont {Lei}, \citenamefont
  {Dong},\ and\ \citenamefont {Wang}}]{Wang2021PhysRevB}%
  \BibitemOpen
  \bibfield  {author} {\bibinfo {author} {\bibfnamefont {Z.~X.}\ \bibnamefont
  {Wang}}, \bibinfo {author} {\bibfnamefont {Q.}~\bibnamefont {Wu}}, \bibinfo
  {author} {\bibfnamefont {Q.~W.}\ \bibnamefont {Yin}}, \bibinfo {author}
  {\bibfnamefont {C.~S.}\ \bibnamefont {Gong}}, \bibinfo {author}
  {\bibfnamefont {Z.~J.}\ \bibnamefont {Tu}}, \bibinfo {author} {\bibfnamefont
  {T.}~\bibnamefont {Lin}}, \bibinfo {author} {\bibfnamefont {Q.~M.}\
  \bibnamefont {Liu}}, \bibinfo {author} {\bibfnamefont {L.~Y.}\ \bibnamefont
  {Shi}}, \bibinfo {author} {\bibfnamefont {S.~J.}\ \bibnamefont {Zhang}},
  \bibinfo {author} {\bibfnamefont {D.}~\bibnamefont {Wu}}, \bibinfo {author}
  {\bibfnamefont {H.~C.}\ \bibnamefont {Lei}}, \bibinfo {author} {\bibfnamefont
  {T.}~\bibnamefont {Dong}},\ and\ \bibinfo {author} {\bibfnamefont {N.~L.}\
  \bibnamefont {Wang}},\ }\bibfield  {title} {\bibinfo {title} {{Unconventional
  charge density wave and photoinduced lattice symmetry change in the kagome
  metal ${\mathrm{CsV}}_{3}{\mathrm{Sb}}_{5}$ probed by time-resolved
  spectroscopy}},\ }\href {https://doi.org/10.1103/PhysRevB.104.165110}
  {\bibfield  {journal} {\bibinfo  {journal} {Phys. Rev. B}\ }\textbf {\bibinfo
  {volume} {104}},\ \bibinfo {pages} {165110} (\bibinfo {year}
  {2021})}\BibitemShut {NoStop}%
\bibitem [{\citenamefont {Ratcliff}\ \emph {et~al.}(2021)\citenamefont
  {Ratcliff}, \citenamefont {Hallett}, \citenamefont {Ortiz}, \citenamefont
  {Wilson},\ and\ \citenamefont {Harter}}]{Ratcliff2021PhysRevMaterials}%
  \BibitemOpen
  \bibfield  {author} {\bibinfo {author} {\bibfnamefont {N.}~\bibnamefont
  {Ratcliff}}, \bibinfo {author} {\bibfnamefont {L.}~\bibnamefont {Hallett}},
  \bibinfo {author} {\bibfnamefont {B.~R.}\ \bibnamefont {Ortiz}}, \bibinfo
  {author} {\bibfnamefont {S.~D.}\ \bibnamefont {Wilson}},\ and\ \bibinfo
  {author} {\bibfnamefont {J.~W.}\ \bibnamefont {Harter}},\ }\bibfield  {title}
  {\bibinfo {title} {{Coherent phonon spectroscopy and interlayer modulation of
  charge density wave order in the kagome metal
  ${\mathrm{CsV}}_{3}{\mathrm{Sb}}_{5}$}},\ }\href
  {https://doi.org/10.1103/PhysRevMaterials.5.L111801} {\bibfield  {journal}
  {\bibinfo  {journal} {Phys. Rev. Materials}\ }\textbf {\bibinfo {volume}
  {5}},\ \bibinfo {pages} {L111801} (\bibinfo {year} {2021})}\BibitemShut
  {NoStop}%
\bibitem [{\citenamefont {Nie}\ \emph {et~al.}(2022)\citenamefont {Nie},
  \citenamefont {Sun}, \citenamefont {Ma}, \citenamefont {Song}, \citenamefont
  {Zheng}, \citenamefont {Liang}, \citenamefont {Wu}, \citenamefont {Yu},
  \citenamefont {Li}, \citenamefont {Shan}, \citenamefont {Zhao}, \citenamefont
  {Li}, \citenamefont {Kang}, \citenamefont {Wu}, \citenamefont {Zhou},
  \citenamefont {Liu}, \citenamefont {Xiang}, \citenamefont {Ying},
  \citenamefont {Wang}, \citenamefont {Wu},\ and\ \citenamefont
  {Chen}}]{Nie2022nature}%
  \BibitemOpen
  \bibfield  {author} {\bibinfo {author} {\bibfnamefont {L.}~\bibnamefont
  {Nie}}, \bibinfo {author} {\bibfnamefont {K.}~\bibnamefont {Sun}}, \bibinfo
  {author} {\bibfnamefont {W.}~\bibnamefont {Ma}}, \bibinfo {author}
  {\bibfnamefont {D.}~\bibnamefont {Song}}, \bibinfo {author} {\bibfnamefont
  {L.}~\bibnamefont {Zheng}}, \bibinfo {author} {\bibfnamefont
  {Z.}~\bibnamefont {Liang}}, \bibinfo {author} {\bibfnamefont
  {P.}~\bibnamefont {Wu}}, \bibinfo {author} {\bibfnamefont {F.}~\bibnamefont
  {Yu}}, \bibinfo {author} {\bibfnamefont {J.}~\bibnamefont {Li}}, \bibinfo
  {author} {\bibfnamefont {M.}~\bibnamefont {Shan}}, \bibinfo {author}
  {\bibfnamefont {D.}~\bibnamefont {Zhao}}, \bibinfo {author} {\bibfnamefont
  {S.}~\bibnamefont {Li}}, \bibinfo {author} {\bibfnamefont {B.}~\bibnamefont
  {Kang}}, \bibinfo {author} {\bibfnamefont {Z.}~\bibnamefont {Wu}}, \bibinfo
  {author} {\bibfnamefont {Y.}~\bibnamefont {Zhou}}, \bibinfo {author}
  {\bibfnamefont {K.}~\bibnamefont {Liu}}, \bibinfo {author} {\bibfnamefont
  {Z.}~\bibnamefont {Xiang}}, \bibinfo {author} {\bibfnamefont
  {J.}~\bibnamefont {Ying}}, \bibinfo {author} {\bibfnamefont {Z.}~\bibnamefont
  {Wang}}, \bibinfo {author} {\bibfnamefont {T.}~\bibnamefont {Wu}},\ and\
  \bibinfo {author} {\bibfnamefont {X.}~\bibnamefont {Chen}},\ }\bibfield
  {title} {\bibinfo {title} {Charge-density-wave-driven electronic nematicity
  in a kagome superconductor},\ }\href
  {https://doi.org/10.1038/s41586-022-04493-8} {\bibfield  {journal} {\bibinfo
  {journal} {Nature}\ }\textbf {\bibinfo {volume} {604}},\ \bibinfo {pages}
  {59} (\bibinfo {year} {2022})}\BibitemShut {NoStop}%
\bibitem [{\citenamefont {Xu}\ \emph {et~al.}(2020{\natexlab{b}})\citenamefont
  {Xu}, \citenamefont {Li}, \citenamefont {Ding}, \citenamefont {Chen},
  \citenamefont {Sakai}, \citenamefont {Fauqu\'e}, \citenamefont {Nakatsuji},
  \citenamefont {Zhu},\ and\ \citenamefont {Behnia}}]{Xu2020PhysRevB}%
  \BibitemOpen
  \bibfield  {author} {\bibinfo {author} {\bibfnamefont {L.}~\bibnamefont
  {Xu}}, \bibinfo {author} {\bibfnamefont {X.}~\bibnamefont {Li}}, \bibinfo
  {author} {\bibfnamefont {L.}~\bibnamefont {Ding}}, \bibinfo {author}
  {\bibfnamefont {T.}~\bibnamefont {Chen}}, \bibinfo {author} {\bibfnamefont
  {A.}~\bibnamefont {Sakai}}, \bibinfo {author} {\bibfnamefont
  {B.}~\bibnamefont {Fauqu\'e}}, \bibinfo {author} {\bibfnamefont
  {S.}~\bibnamefont {Nakatsuji}}, \bibinfo {author} {\bibfnamefont
  {Z.}~\bibnamefont {Zhu}},\ and\ \bibinfo {author} {\bibfnamefont
  {K.}~\bibnamefont {Behnia}},\ }\bibfield  {title} {\bibinfo {title}
  {{Anomalous transverse response of ${\mathrm{Co}}_{2}\mathrm{MnGa}$ and
  universality of the room-temperature
  ${\ensuremath{\alpha}}_{ij}^{A}/{\ensuremath{\sigma}}_{ij}^{A}$ ratio across
  topological magnets}},\ }\href {https://doi.org/10.1103/PhysRevB.101.180404}
  {\bibfield  {journal} {\bibinfo  {journal} {Phys. Rev. B}\ }\textbf {\bibinfo
  {volume} {101}},\ \bibinfo {pages} {180404} (\bibinfo {year}
  {2020}{\natexlab{b}})}\BibitemShut {NoStop}%
\bibitem [{\citenamefont {Hou}\ \emph {et~al.}(2015)\citenamefont {Hou},
  \citenamefont {Su}, \citenamefont {Tian}, \citenamefont {Jin}, \citenamefont
  {Yang},\ and\ \citenamefont {Niu}}]{Hou2015PRL}%
  \BibitemOpen
  \bibfield  {author} {\bibinfo {author} {\bibfnamefont {D.}~\bibnamefont
  {Hou}}, \bibinfo {author} {\bibfnamefont {G.}~\bibnamefont {Su}}, \bibinfo
  {author} {\bibfnamefont {Y.}~\bibnamefont {Tian}}, \bibinfo {author}
  {\bibfnamefont {X.}~\bibnamefont {Jin}}, \bibinfo {author} {\bibfnamefont
  {S.~A.}\ \bibnamefont {Yang}},\ and\ \bibinfo {author} {\bibfnamefont
  {Q.}~\bibnamefont {Niu}},\ }\bibfield  {title} {\bibinfo {title}
  {{Multivariable Scaling for the Anomalous Hall Effect}},\ }\href
  {https://doi.org/10.1103/PhysRevLett.114.217203} {\bibfield  {journal}
  {\bibinfo  {journal} {Phys. Rev. Lett.}\ }\textbf {\bibinfo {volume} {114}},\
  \bibinfo {pages} {217203} (\bibinfo {year} {2015})}\BibitemShut {NoStop}%
\bibitem [{\citenamefont {Gan}\ \emph {et~al.}(2021)\citenamefont {Gan},
  \citenamefont {Xia}, \citenamefont {Zhang}, \citenamefont {Yang},
  \citenamefont {Mi}, \citenamefont {Wang}, \citenamefont {Chai}, \citenamefont
  {Guo}, \citenamefont {Zhou},\ and\ \citenamefont
  {He}}]{gan2021magnetoseebeck}%
  \BibitemOpen
  \bibfield  {author} {\bibinfo {author} {\bibfnamefont {Y.}~\bibnamefont
  {Gan}}, \bibinfo {author} {\bibfnamefont {W.}~\bibnamefont {Xia}}, \bibinfo
  {author} {\bibfnamefont {L.}~\bibnamefont {Zhang}}, \bibinfo {author}
  {\bibfnamefont {K.}~\bibnamefont {Yang}}, \bibinfo {author} {\bibfnamefont
  {X.}~\bibnamefont {Mi}}, \bibinfo {author} {\bibfnamefont {A.}~\bibnamefont
  {Wang}}, \bibinfo {author} {\bibfnamefont {Y.}~\bibnamefont {Chai}}, \bibinfo
  {author} {\bibfnamefont {Y.}~\bibnamefont {Guo}}, \bibinfo {author}
  {\bibfnamefont {X.}~\bibnamefont {Zhou}},\ and\ \bibinfo {author}
  {\bibfnamefont {M.}~\bibnamefont {He}},\ }\bibfield  {title} {\bibinfo
  {title} {{Magneto-Seebeck effect and ambipolar Nernst effect in the
  ${\mathrm{CsV}}_{3}{\mathrm{Sb}}_{5}$ superconductor}},\ }\href
  {https://doi.org/10.1103/PhysRevB.104.L180508} {\bibfield  {journal}
  {\bibinfo  {journal} {Phys. Rev. B}\ }\textbf {\bibinfo {volume} {104}},\
  \bibinfo {pages} {L180508} (\bibinfo {year} {2021})}\BibitemShut {NoStop}%
\bibitem [{\citenamefont {Long}(1967)}]{LONG1967PLA}%
  \BibitemOpen
  \bibfield  {author} {\bibinfo {author} {\bibfnamefont {J.}~\bibnamefont
  {Long}},\ }\bibfield  {title} {\bibinfo {title} {Phonon drag effects in
  tungsten below 2.6°{K}},\ }\href
  {https://doi.org/https://doi.org/10.1016/0375-9601(67)90470-7} {\bibfield
  {journal} {\bibinfo  {journal} {Physics Letters A}\ }\textbf {\bibinfo
  {volume} {25}},\ \bibinfo {pages} {677} (\bibinfo {year} {1967})}\BibitemShut
  {NoStop}%
\bibitem [{\citenamefont {Bel}\ \emph {et~al.}(2003)\citenamefont {Bel},
  \citenamefont {Behnia},\ and\ \citenamefont {Berger}}]{NbSe22003}%
  \BibitemOpen
  \bibfield  {author} {\bibinfo {author} {\bibfnamefont {R.}~\bibnamefont
  {Bel}}, \bibinfo {author} {\bibfnamefont {K.}~\bibnamefont {Behnia}},\ and\
  \bibinfo {author} {\bibfnamefont {H.}~\bibnamefont {Berger}},\ }\bibfield
  {title} {\bibinfo {title} {{Ambipolar Nernst effect in NbSe$_{2}$}},\ }\href
  {https://doi.org/10.1103/PhysRevLett.91.066602} {\bibfield  {journal}
  {\bibinfo  {journal} {Phys. Rev. Lett.}\ }\textbf {\bibinfo {volume} {91}},\
  \bibinfo {pages} {066602} (\bibinfo {year} {2003})}\BibitemShut {NoStop}%
\bibitem [{\citenamefont {{Zengwei Zhu and Huan Yang and Aritra Banerjee and
  Liam Malone and Beno{\^{\i}}t Fauqu{\'{e}} and Kamran
  Behnia}}(2011)}]{Zhu2011JOP}%
  \BibitemOpen
  \bibfield  {author} {\bibinfo {author} {\bibnamefont {{Zengwei Zhu and Huan
  Yang and Aritra Banerjee and Liam Malone and Beno{\^{\i}}t Fauqu{\'{e}} and
  Kamran Behnia}}},\ }\bibfield  {title} {\bibinfo {title} {{Nernst quantum
  oscillations in bulk semi-metals}},\ }\href
  {https://doi.org/10.1088/0953-8984/23/9/094204} {\bibfield  {journal}
  {\bibinfo  {journal} {Journal of Physics: Condensed Matter}\ }\textbf
  {\bibinfo {volume} {23}},\ \bibinfo {pages} {094204} (\bibinfo {year}
  {2011})}\BibitemShut {NoStop}%
\bibitem [{\citenamefont {Li}\ \emph {et~al.}(2021{\natexlab{b}})\citenamefont
  {Li}, \citenamefont {Wu}, \citenamefont {Liu}, \citenamefont {Polley},
  \citenamefont {Guo}, \citenamefont {Wang}, \citenamefont {Han}, \citenamefont
  {Dendzik}, \citenamefont {Berntsen}, \citenamefont {Thiagarajan},
  \citenamefont {Shi}, \citenamefont {Schnyder},\ and\ \citenamefont
  {Tjernberg}}]{Congli2021spectroscopic}%
  \BibitemOpen
  \bibfield  {author} {\bibinfo {author} {\bibfnamefont {C.}~\bibnamefont
  {Li}}, \bibinfo {author} {\bibfnamefont {X.}~\bibnamefont {Wu}}, \bibinfo
  {author} {\bibfnamefont {H.}~\bibnamefont {Liu}}, \bibinfo {author}
  {\bibfnamefont {C.}~\bibnamefont {Polley}}, \bibinfo {author} {\bibfnamefont
  {Q.}~\bibnamefont {Guo}}, \bibinfo {author} {\bibfnamefont {Y.}~\bibnamefont
  {Wang}}, \bibinfo {author} {\bibfnamefont {X.}~\bibnamefont {Han}}, \bibinfo
  {author} {\bibfnamefont {M.}~\bibnamefont {Dendzik}}, \bibinfo {author}
  {\bibfnamefont {M.~H.}\ \bibnamefont {Berntsen}}, \bibinfo {author}
  {\bibfnamefont {B.}~\bibnamefont {Thiagarajan}}, \bibinfo {author}
  {\bibfnamefont {Y.}~\bibnamefont {Shi}}, \bibinfo {author} {\bibfnamefont
  {A.~P.}\ \bibnamefont {Schnyder}},\ and\ \bibinfo {author} {\bibfnamefont
  {O.}~\bibnamefont {Tjernberg}},\ }\href@noop {} {\bibinfo {title}
  {{Spectroscopic Evidence for a Three-Dimensional Charge Density Wave in
  Kagome Superconductor CsV$_3$Sb$_5$}}} (\bibinfo {year}
  {2021}{\natexlab{b}}),\ \Eprint {https://arxiv.org/abs/2112.06565}
  {arXiv:2112.06565 [cond-mat.str-el]} \BibitemShut {NoStop}%
\bibitem [{\citenamefont {Nakayama}\ \emph {et~al.}(2021)\citenamefont
  {Nakayama}, \citenamefont {Li}, \citenamefont {Kato}, \citenamefont {Liu},
  \citenamefont {Wang}, \citenamefont {Takahashi}, \citenamefont {Yao},\ and\
  \citenamefont {Sato}}]{Nakayama2021PRB}%
  \BibitemOpen
  \bibfield  {author} {\bibinfo {author} {\bibfnamefont {K.}~\bibnamefont
  {Nakayama}}, \bibinfo {author} {\bibfnamefont {Y.}~\bibnamefont {Li}},
  \bibinfo {author} {\bibfnamefont {T.}~\bibnamefont {Kato}}, \bibinfo {author}
  {\bibfnamefont {M.}~\bibnamefont {Liu}}, \bibinfo {author} {\bibfnamefont
  {Z.}~\bibnamefont {Wang}}, \bibinfo {author} {\bibfnamefont {T.}~\bibnamefont
  {Takahashi}}, \bibinfo {author} {\bibfnamefont {Y.}~\bibnamefont {Yao}},\
  and\ \bibinfo {author} {\bibfnamefont {T.}~\bibnamefont {Sato}},\ }\bibfield
  {title} {\bibinfo {title} {{Multiple energy scales and anisotropic energy gap
  in the charge-density-wave phase of the kagome superconductor
  ${\mathrm{CsV}}_{3}{\mathrm{Sb}}_{5}$}},\ }\href
  {https://doi.org/10.1103/PhysRevB.104.L161112} {\bibfield  {journal}
  {\bibinfo  {journal} {Phys. Rev. B}\ }\textbf {\bibinfo {volume} {104}},\
  \bibinfo {pages} {L161112} (\bibinfo {year} {2021})}\BibitemShut {NoStop}%
\bibitem [{\citenamefont {Lou}\ \emph {et~al.}(2022)\citenamefont {Lou},
  \citenamefont {Fedorov}, \citenamefont {Yin}, \citenamefont {Kuibarov},
  \citenamefont {Tu}, \citenamefont {Gong}, \citenamefont {Schwier},
  \citenamefont {B\"uchner}, \citenamefont {Lei},\ and\ \citenamefont
  {Borisenko}}]{Lou2022PRL}%
  \BibitemOpen
  \bibfield  {author} {\bibinfo {author} {\bibfnamefont {R.}~\bibnamefont
  {Lou}}, \bibinfo {author} {\bibfnamefont {A.}~\bibnamefont {Fedorov}},
  \bibinfo {author} {\bibfnamefont {Q.}~\bibnamefont {Yin}}, \bibinfo {author}
  {\bibfnamefont {A.}~\bibnamefont {Kuibarov}}, \bibinfo {author}
  {\bibfnamefont {Z.}~\bibnamefont {Tu}}, \bibinfo {author} {\bibfnamefont
  {C.}~\bibnamefont {Gong}}, \bibinfo {author} {\bibfnamefont {E.~F.}\
  \bibnamefont {Schwier}}, \bibinfo {author} {\bibfnamefont {B.}~\bibnamefont
  {B\"uchner}}, \bibinfo {author} {\bibfnamefont {H.}~\bibnamefont {Lei}},\
  and\ \bibinfo {author} {\bibfnamefont {S.}~\bibnamefont {Borisenko}},\
  }\bibfield  {title} {\bibinfo {title} {{Charge-Density-Wave-Induced
  Peak-Dip-Hump Structure and the Multiband Superconductivity in a Kagome
  Superconductor ${\mathrm{CsV}}_{3}{\mathrm{Sb}}_{5}$}},\ }\href
  {https://doi.org/10.1103/PhysRevLett.128.036402} {\bibfield  {journal}
  {\bibinfo  {journal} {Phys. Rev. Lett.}\ }\textbf {\bibinfo {volume} {128}},\
  \bibinfo {pages} {036402} (\bibinfo {year} {2022})}\BibitemShut {NoStop}%
\bibitem [{\citenamefont {Liu}\ \emph {et~al.}(2021)\citenamefont {Liu},
  \citenamefont {Zhao}, \citenamefont {Yin}, \citenamefont {Gong},
  \citenamefont {Tu}, \citenamefont {Li}, \citenamefont {Song}, \citenamefont
  {Liu}, \citenamefont {Shen}, \citenamefont {Huang}, \citenamefont {Liu},
  \citenamefont {Lei},\ and\ \citenamefont {Wang}}]{Liu2021PRX}%
  \BibitemOpen
  \bibfield  {author} {\bibinfo {author} {\bibfnamefont {Z.}~\bibnamefont
  {Liu}}, \bibinfo {author} {\bibfnamefont {N.}~\bibnamefont {Zhao}}, \bibinfo
  {author} {\bibfnamefont {Q.}~\bibnamefont {Yin}}, \bibinfo {author}
  {\bibfnamefont {C.}~\bibnamefont {Gong}}, \bibinfo {author} {\bibfnamefont
  {Z.}~\bibnamefont {Tu}}, \bibinfo {author} {\bibfnamefont {M.}~\bibnamefont
  {Li}}, \bibinfo {author} {\bibfnamefont {W.}~\bibnamefont {Song}}, \bibinfo
  {author} {\bibfnamefont {Z.}~\bibnamefont {Liu}}, \bibinfo {author}
  {\bibfnamefont {D.}~\bibnamefont {Shen}}, \bibinfo {author} {\bibfnamefont
  {Y.}~\bibnamefont {Huang}}, \bibinfo {author} {\bibfnamefont
  {K.}~\bibnamefont {Liu}}, \bibinfo {author} {\bibfnamefont {H.}~\bibnamefont
  {Lei}},\ and\ \bibinfo {author} {\bibfnamefont {S.}~\bibnamefont {Wang}},\
  }\bibfield  {title} {\bibinfo {title} {{Charge-Density-Wave-Induced Bands
  Renormalization and Energy Gaps in a Kagome Superconductor
  ${\mathrm{RbV}}_{3}{\mathrm{Sb}}_{5}$}},\ }\href
  {https://doi.org/10.1103/PhysRevX.11.041010} {\bibfield  {journal} {\bibinfo
  {journal} {Phys. Rev. X}\ }\textbf {\bibinfo {volume} {11}},\ \bibinfo
  {pages} {041010} (\bibinfo {year} {2021})}\BibitemShut {NoStop}%
\bibitem [{\citenamefont {McCormick}\ \emph {et~al.}(2017)\citenamefont
  {McCormick}, \citenamefont {McKay},\ and\ \citenamefont
  {Trivedi}}]{McCormick2017prb}%
  \BibitemOpen
  \bibfield  {author} {\bibinfo {author} {\bibfnamefont {T.~M.}\ \bibnamefont
  {McCormick}}, \bibinfo {author} {\bibfnamefont {R.~C.}\ \bibnamefont
  {McKay}},\ and\ \bibinfo {author} {\bibfnamefont {N.}~\bibnamefont
  {Trivedi}},\ }\bibfield  {title} {\bibinfo {title} {Semiclassical theory of
  anomalous transport in type-ii topological weyl semimetals},\ }\href
  {https://doi.org/10.1103/PhysRevB.96.235116} {\bibfield  {journal} {\bibinfo
  {journal} {Phys. Rev. B}\ }\textbf {\bibinfo {volume} {96}},\ \bibinfo
  {pages} {235116} (\bibinfo {year} {2017})}\BibitemShut {NoStop}%
\bibitem [{\citenamefont {{Wang}}\ \emph {et~al.}(2021)\citenamefont {{Wang}},
  \citenamefont {{Boyack}},\ and\ \citenamefont {{Levin}}}]{Wang2021arxiv}%
  \BibitemOpen
  \bibfield  {author} {\bibinfo {author} {\bibfnamefont {Z.}~\bibnamefont
  {{Wang}}}, \bibinfo {author} {\bibfnamefont {R.}~\bibnamefont {{Boyack}}},\
  and\ \bibinfo {author} {\bibfnamefont {K.}~\bibnamefont {{Levin}}},\
  }\bibfield  {title} {\bibinfo {title} {{Heat-bath approach to anomalous
  thermal transport: effects of inelastic scattering}},\ }\href@noop {}
  {\bibfield  {journal} {\bibinfo  {journal} {arXiv e-prints}\ ,\ \bibinfo
  {eid} {arXiv:2112.13148}} (\bibinfo {year} {2021})},\ \Eprint
  {https://arxiv.org/abs/2112.13148} {arXiv:2112.13148 [cond-mat.mes-hall]}
  \BibitemShut {NoStop}%
\bibitem [{\citenamefont {Ding}\ \emph {et~al.}(2021)\citenamefont {Ding},
  \citenamefont {Koo}, \citenamefont {Yi}, \citenamefont {Xu}, \citenamefont
  {Zuo}, \citenamefont {Yang}, \citenamefont {Shi}, \citenamefont {Yan},
  \citenamefont {Behnia},\ and\ \citenamefont {Zhu}}]{Linchao2021JPS}%
  \BibitemOpen
  \bibfield  {author} {\bibinfo {author} {\bibfnamefont {L.}~\bibnamefont
  {Ding}}, \bibinfo {author} {\bibfnamefont {J.}~\bibnamefont {Koo}}, \bibinfo
  {author} {\bibfnamefont {C.}~\bibnamefont {Yi}}, \bibinfo {author}
  {\bibfnamefont {L.}~\bibnamefont {Xu}}, \bibinfo {author} {\bibfnamefont
  {H.}~\bibnamefont {Zuo}}, \bibinfo {author} {\bibfnamefont {M.}~\bibnamefont
  {Yang}}, \bibinfo {author} {\bibfnamefont {Y.}~\bibnamefont {Shi}}, \bibinfo
  {author} {\bibfnamefont {B.}~\bibnamefont {Yan}}, \bibinfo {author}
  {\bibfnamefont {K.}~\bibnamefont {Behnia}},\ and\ \bibinfo {author}
  {\bibfnamefont {Z.}~\bibnamefont {Zhu}},\ }\bibfield  {title} {\bibinfo
  {title} {{Quantum oscillations, magnetic breakdown and thermal Hall effect in
  Co$_3$Sn$_2$S$_2$}},\ }\href {https://doi.org/10.1088/1361-6463/ac1c2b}
  {\bibfield  {journal} {\bibinfo  {journal} {Journal of Physics D: Applied
  Physics}\ }\textbf {\bibinfo {volume} {54}},\ \bibinfo {pages} {454003}
  (\bibinfo {year} {2021})}\BibitemShut {NoStop}%
\end{thebibliography}
\end{document}